# The Resource Demand of Terawatt-Scale Perovskite Tandem Photovoltaics

**Lukas Wagner[1,*], Jiajia Suo[2], Bowen Yang[2], Dmitry Bogachuk[3], Estelle Gervais[3], Robert Pietzcker[4], Andrea Gassmann[5], Jan Christoph Goldschmidt[1]**

[1] Physics of Solar Energy Conversion Group, Department of Physics, Philipps-University Marburg, Renthof 7, 35037 Marburg, Germany
[2] Department of Chemistry – Ångström Laboratory, Uppsala University, Box 523, SE-75120 Uppsala, Sweden.
[3] Fraunhofer Institute for Solar Energy Systems ISE, Heidenhofstr. 2, 79110 Freiburg, Germany.
[4] Potsdam Institute for Climate Impact Research, P.O. Box 60 12 03, 14412 Potsdam, Germany.
[5] Fraunhofer Research Institution for Materials Recycling and Resource Strategies IWKS, Brentanostr. 2a, 63755 Alzenau, Germany.
* Corresponding author and lead contact (lukas.wagner@physik.uni-marburg.de)

## Summary

**Photovoltaics (PV) is the most important energy conversion technology for cost-efficient climate change mitigation. To reach the international climate goals, the annual PV module production capacity must be expanded to multi-terawatt scale. Economic and resource constraints demand the implementation cost-efficient multi-junction technologies, for which perovskite-based tandem technologies are highly promising. In this work, the resource demand of the emerging perovskite PV technology is investigated, considering two factors of supply criticality, namely mining capacity for minerals, as well as production capacity for synthetic materials. Overall, the expansion of perovskite PV to a multi-terawatt scale may not be limited by material supply if certain materials, especially cesium and indium, can be replaced. Moreover, organic charge transport materials face unresolved scalability challenges. This study demonstrates that, besides the improvement of efficiency and stability, perovskite PV research needs also to be guided by sustainable materials choices and design-for-recycling considerations.**

To limit planetary warming, cross-sectoral rapid transitions are essential. For the energy sector, photovoltaics is the most important energy conversion technology for cost efficient climate change mitigation.[1,2] Therefore, a continued drastic expansion of PV module production is needed, which implies that the PV industry is set to enter multi-terawatt (TW) module production in the coming years.[3] In a recent study, we assessed the resource demand of such multi-TW PV production scenario, applying technological learning models to the current, wafer-based silicon PV market. We found that although the demand for silver and glass as well as energy and associated greenhouse gas emissions are critical, there are no fundamental resource constraints if supply does not decrease and technological learning can be maintained at a high level.[4] Yet, in the study we already outlined that the projected necessary performance increase of PV will surpass the fundamental efficiency limit of silicon PV,[5] which mandates the introduction of novel PV technologies. The candidate with the highest technology readiness level to overcome the single-junction efficiency limit are tandem or multi-junction architectures where multiple solar cells are stacked on top of each other to enable selective harvesting of distinct fractions of the solar spectrum.[6] Because simultaneously, energy demand for solar cell productions needs to reduced considerably, metal halide perovskites have emerged as a highly promising semiconductor for both perovskite/silicon tandem (PST) in the short-term and all-perovskite tandem devices (APT) in the long-term due to their favorable semiconductor properties and processability at ambient conditions with industrially established coating techniques. Today, the efficiency of PST already surpasses the performance limit of silicon devices.[7]

This high potential has led to tremendous research activities on perovskite solar cells (PSC). The vast majority of scientific publications have focused on increasing the power conversion efficiency and stability of laboratory-scale devices (cf. Fig. S1). On par with increasing industry activities, there is also a growing trend to advance the scalability. Fewer studies have been published on questions of sustainability, with a focus on the toxicity of lead or life cycle assessments.[8–10] With the exception of a recent publication focusing on the criticality of indium,[11] the fundamental question whether the materials and processes necessary for all individual layers in perovskite solar cells and all-perovskite tandem solar cells will be available for a TW-scale PV production has not been addressed yet. Instead, it has often been claimed a-priori that perovskite solar cells are made from abundant materials.

In this work, we investigate the hypothesis of resource abundance quantitatively, assess the material demand for a TW-scale perovskite PV production, identify potential supply risks for each material, and derive guidelines for further device optimization and material research. The study is based on a model for future multi-TW perovskite PV production that is coupled to an inventory of the most relevant materials used for PSC production. We find that most materials currently used in perovskite research are likely not linked to a supply risk, although replacements for some commonly used materials need to be found. Two factors of supply criticality are assessed, namely primary production of minerals as well as production capacity for synthetic materials. This approach exceeds the established demand-to-production assessment, highlighting that scaling production from research to industrial level should also be part of resource availability analyses. We identify an urgent need to replace the commonly used metals cesium (used in many perovskite alloys) and indium (employed as transparent electrode). Production of halides and the most promising organic solvents to coat perovskite layers require moderate scale-up. With the exception of PEDOT:PSS, the currently used organic hole transport materials will only be expedient for a multi-TW-scale perovskite PV production if the current material synthesis can be scaled by a factor of ≥10,000.

## Results

**Perovskite PV industry growth scenario.** We base our assessment on a scenario for the growth of global PV capacity in compliance with limiting planetary warming to below 1.5°C, calculated within the REMIND model.[12] Our scenario projects a continuous growth of the PV module production capacity until the end of the century, starting with historical data of global PV module production capacity of approximately 190 $GW_p$ in 2021.[13] In line with PV industry roadmaps,[6] the scenario yields a production capacity of 704 $GW_p$/a in 2030 (cf. Fig. 1a). The 1 $TW_p$/a milestone is surpassed in 2046, and a production of 4 $TW_p$ /a is reached in 2100. These projections fall below historical compound annual growth rates[13] and most other 100% renewable energy scenarios assign similar or higher PV capacity expansion rates by 2050.[1,4] Hence, the estimated material demand should be considered as conservative assessments.

With a market share of 95%, the current global PV module production mainly consists of wafer-based silicon PV technology[13] which experienced a learning rate for power conversion efficiency of 7.9% for each doubling of the cumulative production capacity.[6] To maintain such learning rates, the International Technology Roadmap for Photovoltaic 2022 (ITRPV) forecasts that first tandem devices will enter the mass market between 2024 and 2026, gaining a market share of approximately 5% by 2032.[6] This will likely be realized by perovskite/silicon tandem (PST) devices which possess the highest technological readiness. It is reasonable to assume that, once perovskite PV technology is established with the industrial production of PST, the full economic prospects will be harnessed by replacing wafer-based PST with thin-film all-perovskite tandem solar cells (APT). Consequently, we model the market entry of both PST and APT PV technologies, assuming that a 50% market share will be reached in 2040 for perovskite-silicon tandem PV and in 2050 for all-perovskite tandems (cf. Fig. 1a). Due to the high uncertainty regarding potential market entries, implementations of triple and higher-order junction perovskite PV are not assessed here. Moreover, the focus on tandems does not exclude the market potential for single-junction perovskite PV. For any of these technologies, the material inventory remains the same whereas the material demand scales with the number of junctions. Note that these estimates should not be understood as a market outlook but rather a plausible assessment of the resource requirements if the global demand for PV modules would be fulfilled with perovskite tandem devices.

**Material demand for TW-scale perovskite PV.** We have previously highlighted the criticality in supply of the resource of glass used as PV module substrate and potential back-sheet encapsulation.[4] Furthermore, several studies foresee material constraints and environmental challenges associated with the growing demand of the PV industry for base metals such as Cu, Al, and Fe/steel.[14,15] These materials are used for "peripherical" components like the electrical wiring, mounting and frame and can therefore be regarded as nearly independent of the specific PV technology.

In this work, we exclusively focus on the functional solar cell layers. As depicted in the inset in Fig. 1a, a tandem device consists of a top and bottom solar cell to selectively absorb parts of the solar spectrum, which are connected by an interconnection layer. Perovskite tandem stacks consist of multiple functional thin films comprising the perovskite photoabsorber as well as contact layers and electrodes for charge extraction.

We compiled a comprehensive inventory of the demand of the most frequently used materials and solvents for PSC manufacturing (cf. Supplemental Information – SI section D). Fig. 1b depicts the material demand of 30,170 t/$TW_p$ for a representative all-perovskite tandem stack, illustrating that for TW-scale production, the assessed material demands range in the order of 100 t to 10 kt/a for most materials.

Fig. 1c,d exemplarily shows the annual demand of indium for transparent electrodes and of cesium for perovskites, resulting from the PV growth model.

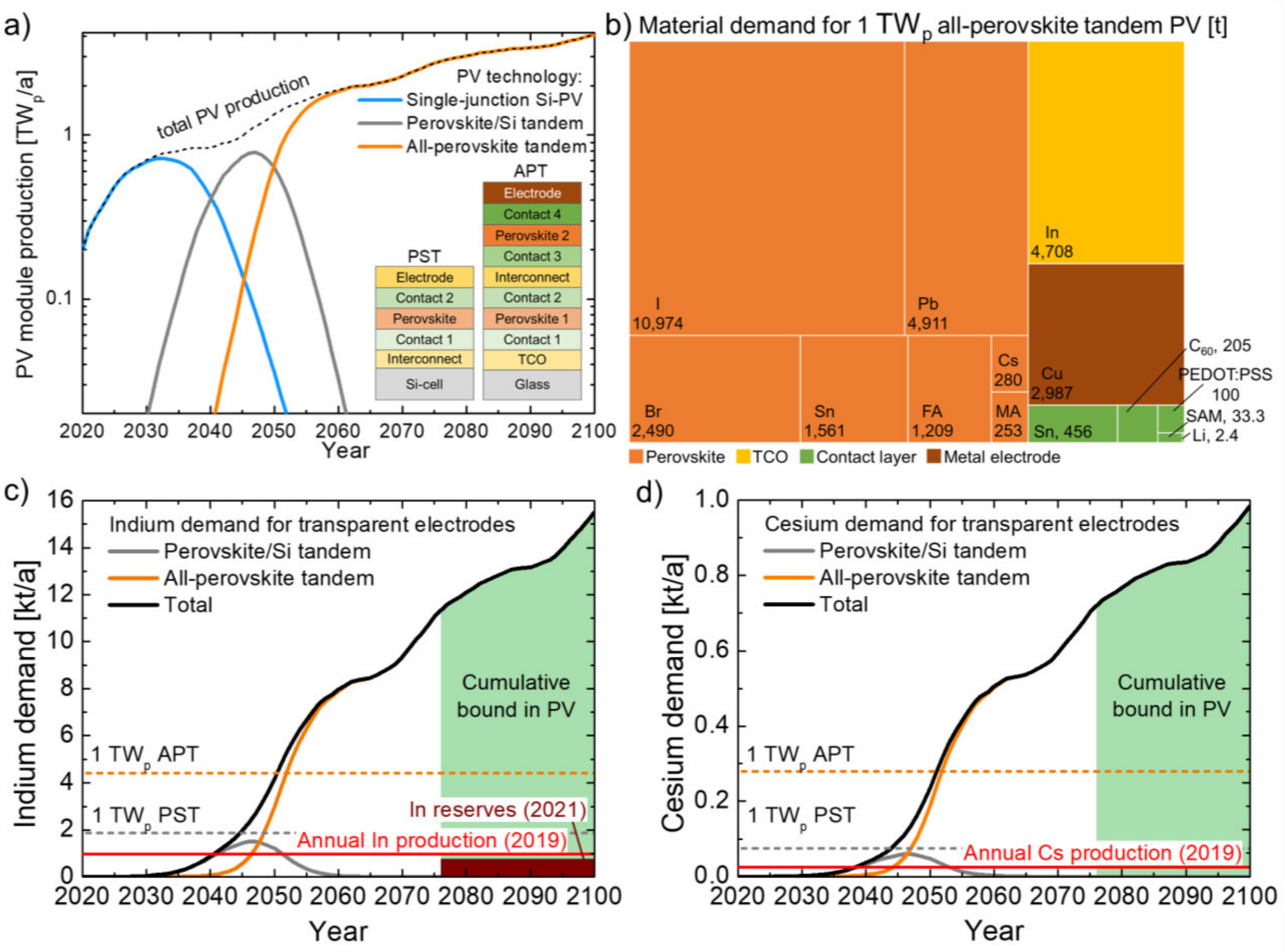

**Fig. 1 | Growth scenarios and material demand for multi-terawatt scale perovskite photovoltaics. a**, Modelled PV module production compatible with an energy infrastructure to reach the 1.5°C climate goal. The scenario considers a technology shift from single-junction silicon PV over perovskite-silicon tandem (PST) peaking in 2047 to all-perovskite tandem (APT) PV. The insets illustrate the layer sequence of typical PST and APT devices (TCO – transparent conductive oxide). **b**, Material demand in metric tons to produce 1 $TW_p$ of a representative all-perovskite tandem (APT) device configuration[16] (cf. Tables S1 & S2). The total material demand of the solar cell layer stacks for 1 $TW_P$ is 30,170 t. **c**,**d**, Annual material demand resulting

from the PV growth model for (**c**) indium used in transparent electrodes, and (**d**) cesium used in perovskites for PST (grey) and APT (orange) devices. Dashed lines indicate the demand for 1 $TW_p$. The solid red line represents the annual In or Cs production in 2019. The green area under the curve represents the highest amount of materials bound in the PV infrastructure in the considered scenarios, assuming module lifetimes of 25 years. For comparison, the area of the dark red box illustrates the currently know In reserves of 18.8 kt.

**Supply criticality of minerals for inorganic materials.** To evaluate and compare the supply criticality, we apply the *demand-production-ratio* (DPR) typically used in supply risk assessments.[17] The ratio relates the material demand for the production of 1 $TW_p$/a PV modules (dashed line in Fig. 1c,d) to current annual primary production (red solid line). While 1 $TW_p$ is a useful functional unit that enables comparability between studies, it is important to note that in our model, this figure will be surpassed by a factor of four by 2100. Fig. 2a summarizes the DPR of the studied inorganic materials (cf. Table S1 for quantitative values). Most materials have a DPR below or in the range of 1%, which can be considered as not critical in supply, assuming that material production will remain at a comparable level. Given the uncertainties associated with the considered long timescales, materials with DPR<10% may also be considered as not critical in supply.

A high risk for supply criticality (cf. discussion in Table 1) is identified for three materials with a DPR above 100%, namely Au (DPR: 195%), and In used in electrodes (DPR: 457% for APT), and especially for Cs used in perovskites (DPR: 1,122%).

Furthermore, a potentially critical range of DPR 10-100% is found for the electrode materials graphite (DPR: 13.1%), and Ag (DPR: 13.2%), In in interconnection layers (DPR: 40% for PST, 30% for APT), and I (DPR: 36.5%) in perovskites. Categorizing these materials as critical in supply requires a more detailed assessment of the current and future primary production. As detailed in Table 1, it is likely that the supply of graphite, and iodine will be sufficiently scalable. Silver consumption remains in the same range as the demand of today's PV industry for 1 $TW_p$ perovskite PV but the demand is exceeded to potentially critical levels for the projected multi-TW production. Indium demand for interconnection layers may become critical.

Using our PV capacity expansion model, we also assess the *bound-reserves-ratio* (BRR). This ratio relates the material assets that are bound in global PV installations to the known reserves of the respective material. In Fig. 1c,d we indicate the bound materials by the shaded green area and the known reserves with the dark red area. Fig. 2b shows a similar behavior of DPR and BRR, but there are also exceptions: due to relatively larger reserves, the BRR is significantly lower than the DPR for graphite (3.3%) and Cs (12.1%). In contrary, limited reserves lead to an even higher BBR compared to the DPR for the Sn in transparent electrode materials (18.0%) and indium (1,809% for electrodes, 118% for interconnection layers). There is also a higher BBR for electrode materials Au (904%), Ag (52.1%), and W (14.1%). The dissipation risk associated with a BBR of 9.3% for 1-nm Au recombination layers is discussed below.

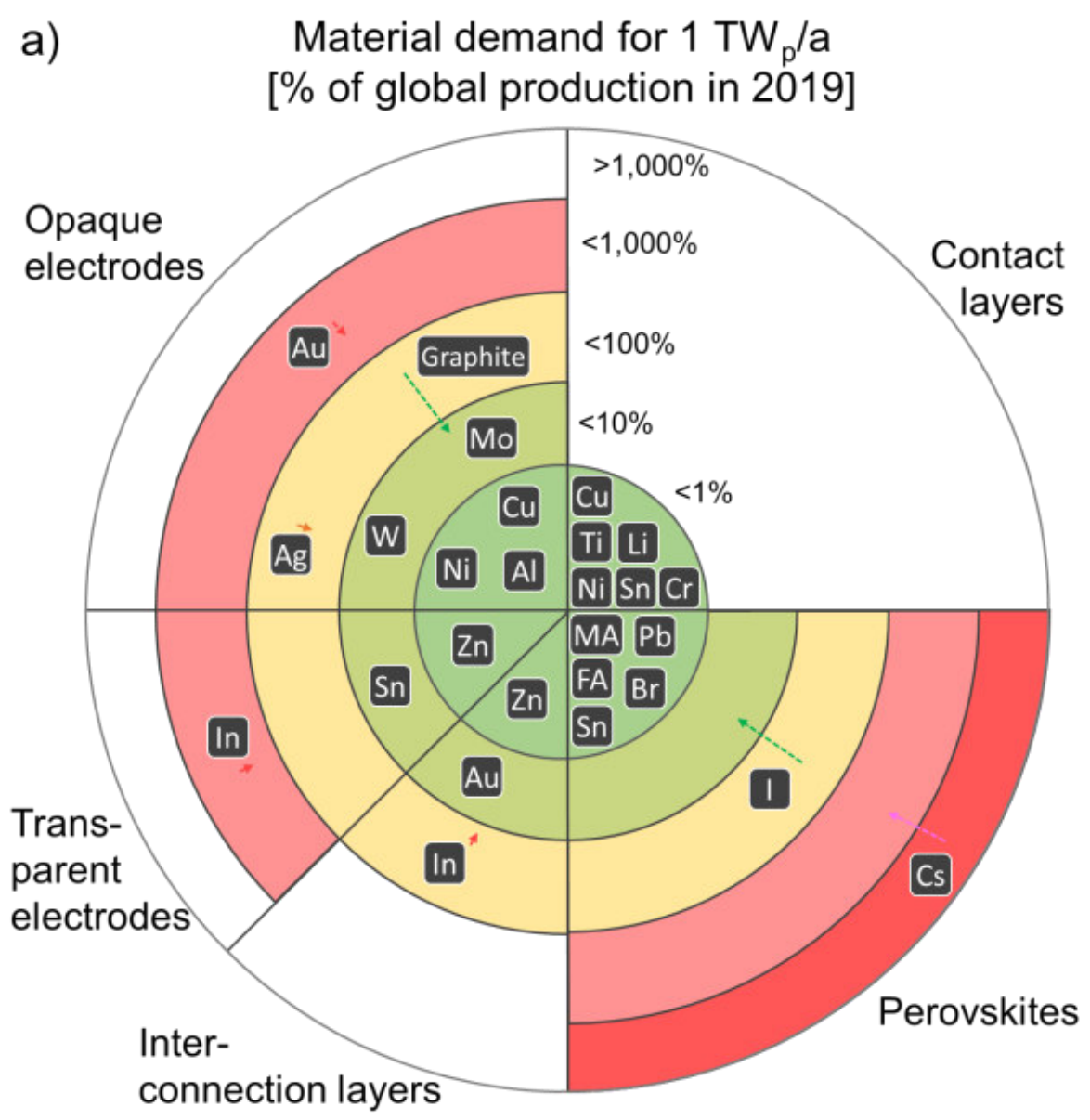

**Fig. 2 | Material demand for TW-scale perovskite PV production and materials bound in multi-TW perovskite PV installations. a**, Supply risk assessment of inorganic materials used for TW-scale perovskite PV. The figure visualizes the demand-production-ratio (DPR) that relates the maximum annual material demand for production of 1 $TW_p$/a perovskite modules to the current annual material mining. For comparability, the organic materials methylammonium (MA) and formamidinium (FA) were been also added. Dashed arrows illustrate potential to reduce the DPR by scaling the production. **b**, Visualization of the bound-reserves-ratio (BRR), relating the materials bound in PV installations with known mining reserves. The BBR of MA, FA, and Br could not be quantified due to unavailable data on reserves. Quantitative values are listed in Table S1. A discussion of the material selection is provided in the SI, section D.

**Supply criticality of synthetic materials.** For advanced synthetic materials like organic materials which mainly consist of carbon and hydrogen atoms, just considering the elemental availability is not sufficient, but also the complexity of the synthesis and current technological readiness of industrial production need to be assessed. The scalability of synthetic materials is discussed in the following. Quantitative values are listed in Table S2 and detailed analysis can be found in the SI Section F.

Methylammonium and formamidinium: Methylammonium (MA) and formamidinium (FA) are essential monovalent cations for perovskites. MA-halides are synthesized by the reaction of halide acids and methylamine, which is synthesized from the highly abundant base chemicals methanol and ammonia (current production – CP>100Mt/a). FA-halides can be synthesized from formamidinium acetate, which is produced by the reaction of cyanamide (CP>1Mt/a) in acetic acid (CP>10Mt/a).[18] Therefore, from the perspective of base material production, MA and FA production for TW-scale perovskite PV is not critical as base material production exceeds the demand by a factor of $10^6$ or $10^5$, respectively.

Organic hole transport materials (HTM): Currently, most organic hole transport materials (HTM) employed in high efficiency PSC such as spiro-OMeTAD (53.9% of all published PSC, cf. Table S5), PTAA (5.1%), or P3HT (2.4%) require Pd-based catalyzers during synthesis and are currently only produced for research. For spiro-OMeTAD, we assessed a global annual production of 7.8 kg/a (cf. Table S9). To satisfy the HTM demand of 100, 1,320, and 1,430 t/a/$TW_p$ production using PTAA, P3HT, or spiro-OMeTAD, respectively, HTM production would need to be expanded drastically (12,800, 168,955, or 182,778 times, respectively; cf. Fig. 3a). With a current production of 227 t of palladium, mainly used for combustion engines, Pd supply represents likely no risk for the scaling in production of these HTM. However, currently, these HTM can only be synthesized in gram-size batches with a low yield (17.8% for spiro-OMeTAD, 43.2% for PTAA), long synthesis durations (>45 h for spiro-OMeTAD, >12 h for PTAA), and complex production routes (7 steps for spiro-OMeTAD, 4 steps for PTAA) requiring various educts, solvents, and acids.[19,20] If these HTM are to be employed for TW-scale perovskite PV production, research on industrial synthesis needs achieve major step-changes.

Self-assembled monolayers (SAM) like 2PACz are the HTM that yield highest efficiencies in tandem PSC.[7,21,22] These materials are currently still expensive to produce. However, they differ from other HTM as only a molecular monolayer of approximately 1 nm thickness is needed. Considering the known synthesis routes, we estimate that the production is principally scalable to the amount of 67 t/a which is necessary for 1 $TW_p$/a perovskite PV. What remains beyond the scope of this work is a detailed assessment of the supply or scalability or substitutability of the necessary educts, especially laboratory-type chemicals.

PEDOT:PSS is an exception among organic HTM as industrial production is already well established for automotive as well as consumer applications. A PEDOT:PSS production of 200 t/a is necessary to fabricate 1 TW/a of perovskite PV, which would correspond to approximately 10,000 t/a of dispersion. Based on our knowledge of the industrial landscape, we estimate that material production at this scale is feasible but might require additional production capacity.

Fullerenes:

The industrial production of fullerenes (e.g., $C_{60}$ or PCBM), which are used as electron transport layer, has already been scaled to an annual production in the multi-ton range. Consequently, although there is a considerable degree of uncertainty about the precise production volume, reaching the demand of 220 t/a/$TW_p$ for all-perovskite tandem solar cells appears to be feasible from a material supply perspective (cf. SI section F for detailed discussion).

$TiO_2$ nanoparticles: Most of the high-efficiency single junction PSC contain a hole-blocking compact layer of titania (c-$TiO_2$) and an electron-selective mesoporous titania (m-$TiO_2$) (cf. Table S5). These materials are produced by different methods with a common precursor of $TiCl_4$ which can be either combined with isopropanol to produce titanium isopropoxide for the deposition of c-$TiO_2$, or used for synthesis of $TiO_2$ nanoparticles by flame spray pyrolysis method,[23] from which m-$TiO_2$ layer is later fabricated. $TiCl_4$ is typically synthesized from the most common Ti-ore ilmenite ($FeTiO_3$) which is mined at a rate of 7.6 Mt/a,[24] corresponding to a DPR of 0.049%. $TiO_2$ particles have been used for paints and inks for decades, highlighting the maturity of $TiO_2$-np industry. Already today, the $TiO_2$-np production of 3 kt[25] surpasses the demand (1.9 kt) for 1 TW/a of all-perovskite tandem solar cells

Solvents: A key appeal of PSC is that most of the functional layers can be dissolved and processed in liquid form. For perovskites absorbers, the most common solvent are blends of dimethylformamide (DMF) and dimethyl sulfoxide (DMSO).[26] Notable alternatives are N-methylpyrrolidone (NMP) and γ-butyrolactone (GBL).[26] Recently, highly volatile solvents with faster evaporation rates have been considered for large scale deposition such as acetonitrile (ACN), ethanol (EtOH), or tetrahydrofuran (THF).[27] For contact layers, most common solvents are toluene (Tol), chlorobenzene (CB), EtOH, isopropyl alcohol (IPA), or $H_2O$. Fig. 3b compares the solvent demand for 1 $TW_p$/a with the current production, demonstrating that there is sufficient solvent production for contact layers whereas a manageable scale-up of up to 20% may be necessary for the additional demand for perovskite absorber deposition. Moreover, as discussed in the next section, the consumption may be drastically decreased with onsite solvent recycling. Therefore, we conclude that solvent consumption reaches volumes that require close consideration, but there will be likely enough supply for TW-scale perovskite PV.

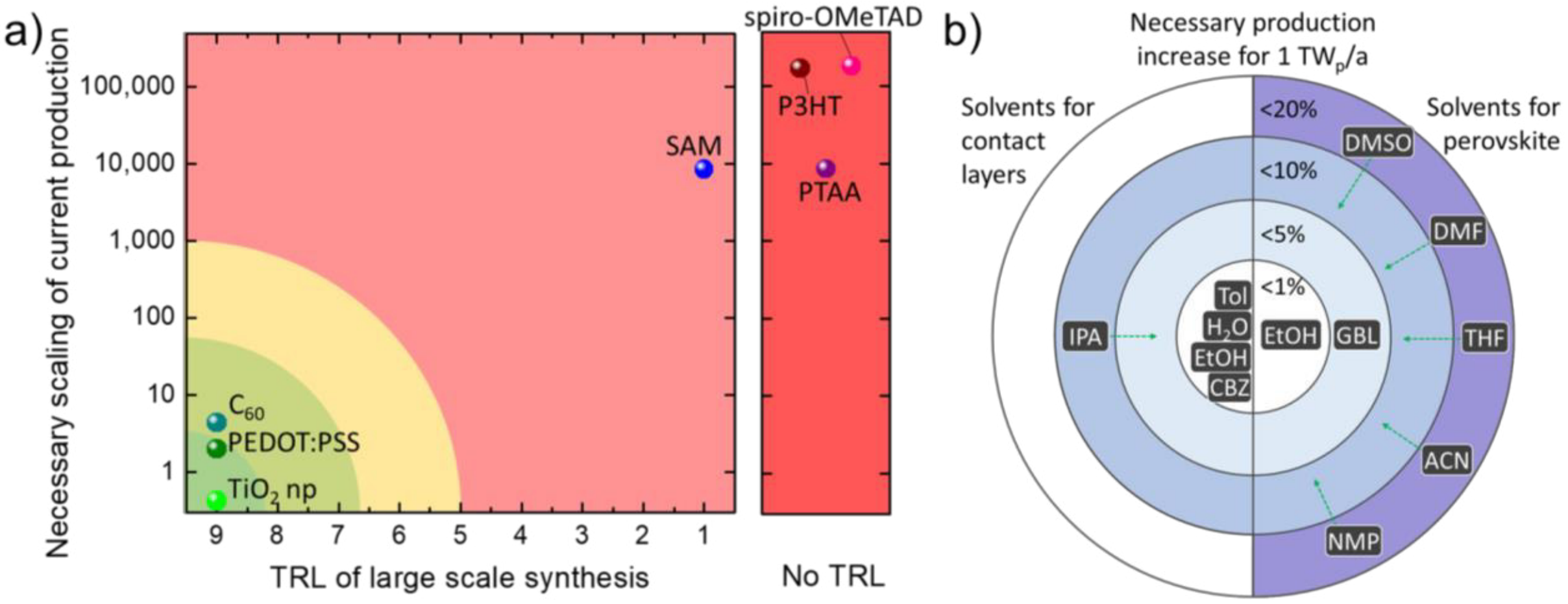

**Fig. 3 | Required increase of production increase for synthetic contact materials and solvents. a**, Necessary scaling of production and increase in technological readiness level (TRL) of the industrial synthesis of synthetic materials for a 1 $TW_p$/a perovskite PV production. No TRL means that no technological concept for industrial production has been formulated yet. **b**, Necessary increment of solvent production for the production of 1 $TW_p$/a perovskite PV. Dashed arrows illustrate potential to reduce consumption by solvent recycling.

## Discussion

**Perovskite PV recycling and material losses.** As material demand keeps increasing strongly, the material feedstock of the PV infrastructure can only to a small part be gained by recycling of discarded PV modules and hence strongly depends on primary material production. Nonetheless, the massive material streams in the range of kilo-tons for individual materials require thorough consideration of dedicated recycling strategies already in early stages of technology development. Therefore, high recycling efficiencies need to be achieved for end-of-life modules and production scrap in all three recycling stages of collection, pre-processing to separate various materials, and end-processing to achieve purified materials (cf. Fig. 4).[28]

Measured by mass, the PV industry is by far the largest semiconductor industry. For recycling this can be advantageous because it provides a large and rather homogeneous material base for effective waste management. End-of-life modules, however, possess a relatively low value-to-weight ratio which makes collection and pre-processing distance sensitive. On the other hand, strongly reduced capital investment costs for perovskite PV module plants open opportunities for decentralized production.[29] This offers the possibility to establish rather local circular economy material networks.

In contrast to established PV technologies,[30] most layers of the PSC stack can be separated and concentrated by low-temperature leaching processes via selective organic solvents, which also opens possibilities for innovative design-for-recycling concepts.[31] This applies especially to inorganic materials. In principle, organic materials can also be recycled from end-of-life modules since they show different solubility in low polar solvents compared to perovskite, but this might be challenging and uneconomic in practice, as these materials will usually have degraded, e.g., by oxidation. As for solvent recycling during the production phase, recovery rates of 71% have been estimated if the solvents are captured in a condenser and purified in a distillation process.[32] Some materials like transparent conductive oxides (TCO) cannot be recovered by leaching. If fluorine-doped tin oxide (FTO) is used as TCO, efficient recycling of FTO-glass should be implemented as the bound-reserves-ratio for Sn in FTO is 18.0%.

Finally, reducing material consumption is not an unconditionally advantageous strategy for more resource efficiency. Dissipation of minute layer thicknesses on large cumulative module areas may pose a serious challenge for the goal of a circular economy. For example, thin 1 nm layers of gold may not be economically recoverable for individual recycling sites and perovskite PV production might globally add up to dissipative gold losses of 64 t/a/$TW_p$. Moreover, the entire all-perovskite tandem infrastructure would bind up to 4.95 kt of Au which corresponds to almost 10% of global Au reserves (BRR = 9.3%). Thus, dissipation losses need to be considered in early stages of research.

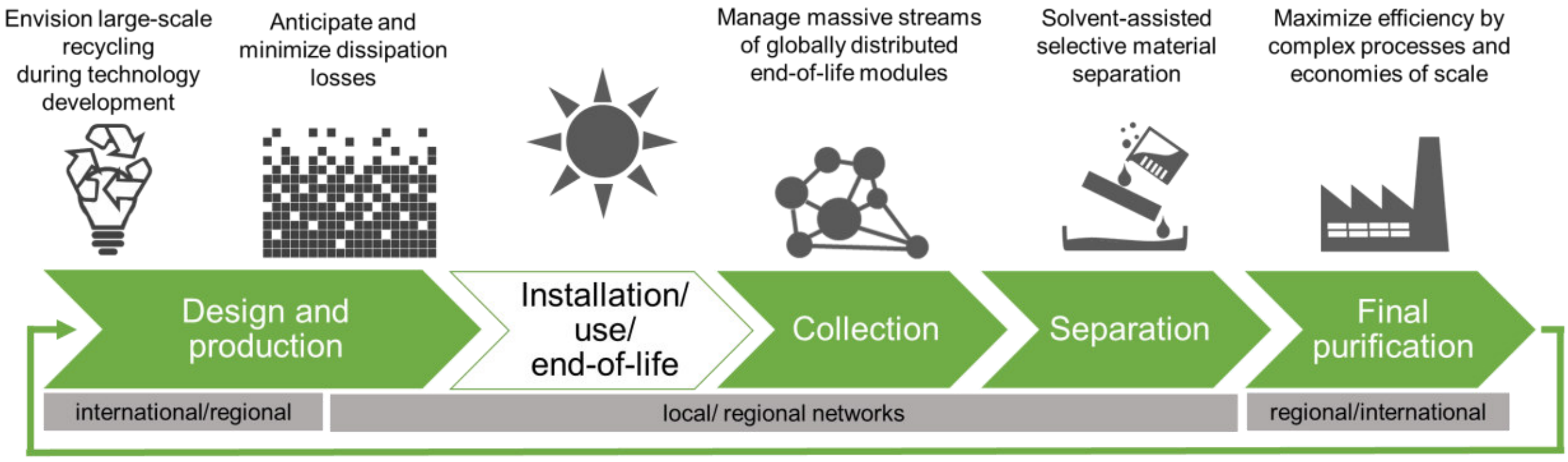

**Fig. 4 | Guidelines for perovskite PV recycling.** Perovskite PV offers promising advantages for innovative recycling strategies if the challenges of massive waste streams are addressed in early stage of technology research**.** End-of-life module recycling is carried out in three stages of module collection, materials separation and final purification. Especially the first two stages can well be embedded in networks of local production, use and recycling, while less mass-intensive final purification may benefit from centralized economies of scale. Solvent assisted material separation is promising to increase the concentration yield supplied to final purification stage. In all stages, dissipation losses must be anticipated and minimized.

**Conclusions for perovskite PV research.** In this study, we assessed the material demand of a multi-terawatt scale perovskite tandem PV infrastructure. By comparing the results with current material production, we draw the following conclusions: multi-terawatt scale perovskite PV is feasible from a material supply perspective; the statement that PSC are made "from abundant materials" is generally justified. However, especially the materials In, Au, and Cs are associated with high supply criticality as the amounts required for 1 TW/a PSC production represent approximately 200%, 460%, and 1,100% of current production, respectively. This is a critical finding as current research activities, driven by the paradigm established by the highest performing devices, are highly focused on device architectures that employ these materials. Indium must be replaced in transport layers, but already the (smaller) demand from its use in interconnection layers is associated with supply risks. Research should focus on stable cesium-free perovskites as Cs will likely not be possible to scale production accordingly. This finding is especially critical for the research direction of completely inorganic perovskites as well as for the use of Cs in stable high bandgap perovskite absorbers. Furthermore, Au and Ag metal electrodes need to be replaced by abundant base metals such as Cu or Al or by graphite.

With the exception of PEDOT:PSS, none of the synthesis routes of organic charge transport materials are currently compliable with industrial large-scale production. Pathways for economic upscaling of production need to be urgently researched. On the other hand, inorganic nanoparticle materials have a high technological maturity and low raw material demand compared to current annual production.

The layer thickness and hence absolute material demand of the active perovskite PV devices stack is approximately 200 times lower than that of established Si-based PV devices. Yet, this work demonstrated that despite using layer thicknesses in the sub-micrometer range, 1 $TW_p$/a perovskite PV production implies annual material demands in the range of kilotons. For the initial build-up of the global PV infrastructure, little of the required material can be withdrawn from existing material stocks within PV, which implies containing or expanding mining activities or the use of others material stocks, e.g., from fossil industries. An important question which is beyond the scope of this work is whether the necessary mining rates can be achieved and what will be the implication for humans, the environment and the planet. Therefore, further works should focus on other aspects of criticality such as supply chain vulnerability or social and ecological aspects. In any case, the large material stocks aggregated in PV modules and the foreseeable massive material waste streams mandate that researchers adapt a resource sensitivity and a "design-for-recycling" thinking already in early stages of technology.

# Experimental procedures

**Expansion scenarios for the PV industry.**

We use the global multi-regional energy-economy-climate model REMIND Version 2.1.0 for our analysis.[12] REMIND is open source and available on GitHub at https://github.com/remindmodel/remind. The technical documentation of the equation structure can be found at https://rse.pik-potsdam.de/doc/remind/2.1.0/. In REMIND, each single region is modelled as a hybrid energy-economy system and is able to interact with the other regions by means of trade. Tradable goods are the exhaustible primary energy carriers coal, oil, gas and uranium, emission permits, and a composite good that represents all other tradeable goods. The economy sector is modelled by a Ramsey-type growth model which maximizes utility, a function of consumption. Labor, capital and end-use energy generate the macroeconomic output, i.e., gross domestic product (GDP). The produced GDP covers the costs of the energy system, the macroeconomic investments, the export of a composite good and consumption.

The energy sector is described with high technological detail. It uses exhaustible and renewable primary energy carriers and converts them to final energy types such as electricity, heat and fuels. Various conversion technologies are available, including technologies with carbon capture and storage (CCS). The model includes cost mark-ups for the fast up-scaling of investments into individual technologies; therefore, a more realistic phasing in and out of technologies is achieved.

The scenario used in this paper follows the default settings of the model, which were also used for scenario S1 in our earlier study.[4] To match the historic standing capacity data in the year 2020, 47 $GW_p$ were added to the REMIND data in each year. Furthermore, the REMIND model delivers the cumulative installed PV capacity in time steps of 5 years until 2060 and in time steps of 10 years until 2100. To obtain data with a yearly resolution, in a first step preliminary installed PV capacity data was estimated for the intermediate years using the compound annual growth rate (CAGR) calculated for the 5- or 10-years time period. Lower CAGR values in a following time period then sometimes lead to a sharp drop in the annual active capacity expansion, when going from one time period to the next. To avoid these artificial oscillations, the data for the annual active capacity expansion was smoothed by a moving average over ±4 years. Finally, by integrating over these values and smoothing with a moving average over ±1 year we obtained the data for the installed PV capacity data without implausible jumps at the transition between time periods, which is used in the present study. While there is a certain discrepancy in the transition from the historic to the projected data, after 2025 the deviation from the REMIND data-points lies below 2.7%. The discrepancy decreases steadily and is below 1% from 2045 on. Respective data for annual active capacity expansion was calculated from the difference in installed PV capacity to the previous year.

To account for the module lifetime $\tau$ of 25 years, the PV production in year $t$, $P^{\tau}(t)$, is calculated from the sum of the active capacity expansion in the same year P(t) and the production from $\tau$ years ago by

$$P^{\tau}(t) = P(t) + P^{\tau}(t-\tau) \tag{1}$$

**Modeling of perovskite PV market shares.** As the PV market is dominated by silicon PV, it is likely that the first tandem PV modules to enter the market will use silicon as bottom-cell.[33] Once perovskite PV technology is industrially established via perovskite/silicon tandems, it is plausible to assume that the market will move towards all-perovskite tandem and multi-junction devices to unravel the full potential of the perovskite technology. The present study focuses on tandem (two junction) cells. Even higher power conversion efficiencies can be reached with a larger number of junctions.[34] However, it is difficult to speculate on the future development of multi-junction perovskite PV such as the market entry and number of junctions. The material demand for devices with more junctions can be interpolated from our assessment, considering that the number of layers will increase. Thereby, it can be assumed that, except for the perovskite layer, the material compositions and layer thicknesses will be not be fundamentally altered.[35,36]

To model the market entry of new perovskite technologies, logistic growth was assumed to the annual production of the new technology $P_{new}$ in year $t$

$$P_{new}(t) = \frac{P_{tot}(t)}{1 + e^{-k\cdot(t-t_0)}} \tag{2}$$

Herein, $P_{tot}(t)$ is the total PV production in year $t$, $k$ describes the growth rate. $t_0$ marks the time when 50% of market share is reached. We assumed that for perovskite-silicon tandem devices, $t_0$ will be reached in 2040, and in 2050 for all-perovskite tandems. A discussion of the assumptions underlying the perovskite PV growth model is detailed in SI Section C.

**Plausibility of assumed growth rate.** In Equation (2), $k$ is a factor that describes the growth rate, which was set to 0.36 for perovskite/silicon tandem PV to be in line with the projections of the International Roadmap for Photovoltaic.[6] Comparing with historic data, this can be considered as a conservative estimate. For comparison, between 2016 and 2020, the growth of the market share of monocrystalline silicon PV to the total PV market could be fit by a logistic growth curve with a growth factor $k$ of 0.6 (cf. Fig. S2). We chose a considerably lower value as monocrystalline silicon PV was already an established technology by 2016. Likewise, a slightly higher growth factor of 0.4 was assumed for all-perovskite tandems to account for the fact that by that time, technological readiness of perovskite PV technologies will be industrially mature, however, only for coating sizes of silicon wafers whereas large scale thin film coating techniques still need to come to maturity.

**Module area determination.** To assess the produced module area, first the annual PV module capacity addition was calculated from the installed capacity estimated by the REMIND model. With the annual module efficiency, projected with an efficiency learning rate of 7.9% for each doubling of the cumulative module production, this yields the annual produced module area. For the reference of 1 $TW_p$/a module production, a module efficiency of 30% was assumed. This is an arbitrary, but representative value (reached by 2053 in our model) which results in a module production of $3.33 \cdot 10^9$ m²/a.

**Materials inventory and layer thicknesses.** A detailed discussion of the selection of considered materials can be found in Supplemental Information SI, Section D. The material demand is governed by the required layer thickness. In specific solar cell stacks, these thicknesses are optimized to yield the optimal device performance. Therefore, it is not possible to assign to a specific material a definite layer thickness. In this study, we considered the layer thicknesses of the latest laboratory efficiency records or of other high performing device stacks to represent a well-optimized system. It is reasonable to assume that, although the layer thickness of future optimized device stacks may vary, the concrete numbers may not differ by orders of magnitudes, which is the core focus of the present supply criticality assessment. The layer thicknesses used for the assessment are listed in Table S6.

**Assessment of solid materials demand.** The material demand of solid material is listed in Table S1. For all-perovskite tandems, the material consumption of contact layers assumes that the respective layer is used twice, once in the top and once in the bottom cell. For inorganic materials, the consumption of the mass of elemental materials per unit area was computed. Therefore, the effective density of each element in the respective material was calculated (e.g., Pb in perovskite). For perovskite blends, material densities were assessed by a linear interpolation between reported values of perovskite compounds.

**Assessment of solvent demand.** For the assessment of the solvent demand of perovskite precursors, first the solvent demand to coat the targeted layer thicknesses was estimated based on precursor concentrations reported in the respective publications referenced in Table S6. Spilling losses during deposition were not considered. For perovskite/silicon tandems, this yields a demand of 1.79ml/m² DMF and 0.446 ml/m² DMSO. The formation of the two perovskite layers in the all-perovskite tandem architecture requires 6.07 ml/m² DMF and 1.03 ml/m² of DMSO, i.e., the total solvent demand is 7.09 ml/m². For a direct comparison between different solvents, we set the solvent demand to 7 ml/m² for all solvents. This is in accordance to the assessment by Vidal et al.[32] who assumed 2.5 ml/m², considering that they studied single-junction devices which contain only one perovskite layer and have lower layer thicknesses than the perovskite layers in all perovskite tandem devices. For all-perovskite tandem devices, two layers – one for each subcell - of each contact layer were assumed in our calculation.
To assess the solvent demand for contact layers, concentrations of 2 mg/ml PTAA in toluene,[16] 0.375 mg/ml SAM in ethanol,[16] 90.9 mg/ml spiro-OMeTAD in chlorobenzene,[37] and 10 mg/mL PCBM in chlorobenzene[38] were assumed, taking representative literature values as a basis. To also include isopropanol in the study, the same solvent demand as estimated for ethanol was assumed.

**Assessment of material supply**. Data on supply and reserves for inorganic materials were obtained from publications of the United States Geological Survey[24] unless otherwise noted (cf. Table S7). SI section F comprises a comprehensive discussion of synthesis routes for synthetic solid materials. Numbers for annual solvent production are listed in Table S11.

**Sensitivity analysis.** A sensitivity analysis for the materials cesium and indium, identified as supply critical in this study, have been carried out in SI Sections H and I, respectively.

## Acknowledgements

The work on this publication was supported by European Union's Horizon Europe Framework Programme for research and innovation under grant agreement no. 101084124 (DIAMOND). J.S. acknowledges the funding from Swiss National Science Foundation for financial support with Project No. 200020_185041.

The authors acknowledge the valuable revisions of Christian Hagelüken (Umicore AG&CoKG), especially concerning the discussion of recycling, and Wilfried Lövenich (Heraeus Deutschland GmbH&CoKG) for valuable contributions to the assessment of PEDOT:PSS supply. The authors thank Nils Langlotz (University of Marburg) for his help with the assessment of solvent demands, Michael Daub (University of Freiburg) for providing data of perovskite densities and Cris C. Tuck (USGS) for valuable feedback on Cs production.

## Author contributions

L.W. and J.C.G. conceived the research question; L.W. assembled the data, conducted the calculation, and coordinated the cooperation; L.W. and R.P. developed the perovskite PV capacity expansion model; L.W. assessed the supply criticality of inorganic compounds; J.S, B.Y, D.B, and L.W. assessed the supply criticality of synthetic material; L.W. conceived the discussion of perovskite PV recycling and material losses; L.W., J.C.G., A.G., and E.G. analyzed the results and assessments of supply criticality; L.W. wrote the manuscript and SI. J.C.G. revised and improved the manuscript and SI. All authors reviewed the final manuscript and SI.

## Competing interest

The authors declare no competing interest.

# Table

**Table 1. Overview of current production and assessment of future supply of the materials graphite, iodine, silver, gold, indium, and cesium.** The materials have been selected due to a high demand-production-ratio (DPR) and bound-reserves-ratio (BRR). The survey is used to assess of the supply criticality for terawatt-scale perovskite PV.

**Graphite**

**Occurrence and production**. Graphite consists of crystalline carbon in the form of stacked graphene sheets of various crystalline forms of which several have been successfully implemented in PSC.[39] Currently, graphite is mostly produced from mining, but can also be fabricated from waste biomass.[40]

**Potential future supply**. Today, around 14% of the global graphite supply is used for anode fabrication in batteries.[41] The International Energy Agency projects a 25-fold increase in global graphite demand by 2040 for the production of batteries for electric vehicles.[42] Despite high projected growth rates of graphite production (6-9%), existing capacities have been assessed to sufficiently satisfy current and future demands.[41] The projected production increase for batteries would lead to a supply 50 times higher than the demand of the perovskite PV industry.

**Assessment of global supply criticality**: low to medium criticality.

**Substitutes.** Conductive metals, conductive synthetic materials.

**Iodine**

**Occurrence and production.** Iodine is currently produced as by-product in sodium nitrate mines or brines occurring in natural gas production.[24]

**Potential future supply.** The current production of I is comparably low, which is reflected in a DPR of 36.5%, with known reserves[24] resulting in a BBR of 15.3%. While currently not economic, resources might be further expanded by the rich resources of the oceans, containing 90 Gt of I at a concentration of 0.06 ppm.[24] Seaweeds enrich iodine to 0.45%.[24] Brines from water desalination may also become attractive.

**Assessment of global supply criticality**: medium criticality.

**Substitutes**. None. Iodine is essential for band-gap optimization.

**Silver and Gold**

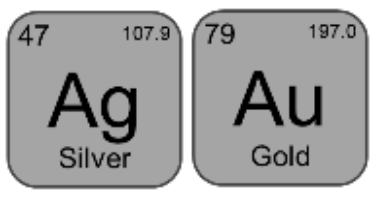

**Occurrence and production.** Au and Ag are some of the oldest metals used by humans. Mining and exploration are accordingly highly mature.

**Silver supply.** Today, the PV industry consumes approximately one tenth of the global silver production.[4] This renders Ag electrodes tolerable for 1 $TW_p$/a perovskite PV scenario, but not for multi-TW production.

**Gold supply.** Considering that Au mining rates might not be able to be increased extensively, gold is not suited as electrode for TW-scale perovskite PV. For completeness, it needs to be added that today, less than 10% of Au mining is used for technological applications and, likewise, large gold stocks are held by central banks. This makes it at least theoretically conceivable that Au is withdrawn from these uses to produce some $TW_p$ of perovskite PV over a limited time. Yet, this appears to be an unlikely scenario as gold electrodes are too costly for competitive PV modules.[43]

**Gold recycling**

The use of nanometer-thin Au as interconnection layers is likely feasible from a supply perspective (DPR: 2.0%). However, it may be connected to high dissipation losses (64 t/$TW_p$) if it is not effectively recycled. In a best case scenario where 1-nm Au is deposited on a module with 2 mm glass/foil module, the Au concentration in the module is approximately 4 g per ton of waste, which is 38 times lower than the Au content in a smart phone (150 g/t), but at the same level as in gold mines.[28] However, as collection and separation has to be considered as cost factor as well, partial recycling of such thin gold layers may only be economic if the gold-layer can be separated in a way that enables Au-concentration during recycling. Eventually, the profitability of Au-recycling will depend on the revenues of other recycled materials.

**Assessment of global supply criticality**: high criticality.

**Substitutes.** Conductive metals such as Cu, Al, Ni. Graphite. For interconnection layers, e.g., heterojunctions from electron and hole transport layers, or transparent conductive oxides.

**Indium**

**Occurrence and production.** Indium is mainly used as transparent conductive oxide in display production, typically as indium tin oxide (ITO - 90-95% of $In_2O_3$ alloyed with 5-10% $SnO_2$).[24,44] The statistical concentration if indium in the earth crust is comparable to Ag. However, in contrast to the more common silver-minerals, there are only twelve known minerals that contain indium.[44] Due to the low concentration in ores, In is therefore only extracted as by-product, typically as a trace constituent from sulfate minerals, mainly zinc-sulfate.[24] Indium mining is hence tightly coupled to the mining rates of base metals, the exclusive extraction of indium is not economic.[45] The demand of In for transparent electrodes exceeds the current production by a factor of 2 to 4.5 for perovskite/silicon tandems (PST) or all-perovskite tandems (APT), respectively. For interconnection layers, the DPR is 40% and 30%, respectively.

**Potential future supply.** Interpolating from reserves of the base metals zinc and copper, the global indium reserves have been estimated to 18.8 kt (15.0 kt for Zn alone).[44,46] This results in a BRR of 1,809% for transparent electrodes and 118% for interconnection layers.

**Assessment of global supply criticality**: high criticality for transparent electrodes, medium to high criticality for interconnection layers.

**Substitutes.** For transparent electrodes, e.g., aluminum doped zinc oxide (AZO), fluorine doped tin oxide (FTO), PEDOT:PSS. For interconnection layers, e.g., heterojunctions from electron and hole transport layers.

**Cesium**

**Occurrence and production.** The statistical occurrence in the earth crust (3 ppm) is comparable to Sn. However, Cs (as well as Rb) belongs to the large-ion lithophile elements. Due to a large ion radius in combination with comparably low ion potential, it is counted as “incompatible element” and crystallizes only in few minerals which renders it highly dispersed.[47] The current Cs production is low and estimated to be 50t/a.[48]

**Potential future supply.** Today, Cs is mainly used for drilling in oil and natural gas production.[24] In a post-fossil world, this share may be used for perovskite PV. However, this potential market shift will not able to offset the high demand for Cs from perovskite PV production as Cs demand for 1 $TW_p$/a perovskite PV requires a 10-fold increase of the current production (DPR: 1,122%). Even if the Cs content of the perovskite is reduced from the assumed 20% to 2%, the demand exceeds the current production (cf. Table S12). Multi-TW scale PV requires even larger increases of Cs mining. Regarding the reserves of below 200 kt,[24] with a BBR of 12.1% a scale-up of production is at least theoretically conceivable.

**Substitutes.** Methylammonium or formamidinium. Cs-free APT have been successfully demonstrated.[49]

## Supplemental information

The Supplemental information (SI) document contains nine sections:

A. Tabulated values of material demands and supply criticalities.
B. Analysis of literature on perovskite PV and sustainability.
C. Further details on the perovskite PV industry growth model.
D. Materials and layer inventory.
E. Production and reserves of minerals.
F. Synthesis routes and estimates of supply for solid synthetic materials.
G. Literature data on global solvent production.
H. Sensitivity analysis of cesium demand.
I. Sensitivity analysis of indium demand.

# Supplementary Information

# The Resource Demand of Terawatt-Scale Perovskite Tandem Photovoltaics

**Lukas Wagner,*[a] Jiajia Suo,[b] Bowen Yang, [b] Dmitry Bogachuk,[c] Estelle Gervais,[c] Robert Pietzcker,[d] Andrea Gassmann[e] and Jan Christoph Goldschmidt[a]**

[a] Physics of Solar Energy Conversion Group, Department of Physics, Philipps-University Marburg, Renthof 7, 35037 Marburg, Germany. E-mail: lukas.wagner@physik.uni-marburg.de
[b] Department of Chemistry – Ångström Laboratory, Uppsala University, Box 523, SE-75120 Uppsala, Sweden.
[c] Fraunhofer Institute for Solar Energy Systems ISE, Heidenhofstr. 2, 79110 Freiburg, Germany.
[d] Potsdam Institute for Climate Impact Research, P.O. Box 60 12 03, 14412 Potsdam, Germany.
[e] Fraunhofer Research Institution for Materials Recycling and Resource Strategies IWKS, Brentanostr. 2a, 63755 Alzenau, Germany.

## Contents

# A. Tabulated values of material demands and supply criticalities

**Table S 1 | Material demand and supply criticalities of materials based on minerals for the perovskite, electrodes, interconnection and inorganic contact layers.** Although they are organic molecules, MA and FA are included here for comparability with other perovskite materials. DPR – demand-production ratio, BRR – bound-reserves-ratio, PST – perovskite/silicon tandem, APT – all-perovskite tandem.

| Material | Layer | Demand for 1 $TW_p$ [t/a/$TW_p$] | | Max. demand[1] [t/a] | DPR [%] | Max. bound material[2] [kt] | BRR [%] |
|---|---|---|---|---|---|---|---|
| **Perovskites** | | **PST** | **APT** | | | | |
| **Cs** | PST-top, APT-bottom | 74.8 | 280 | 985 | 1,122 | 24.2 | 12.1 |
| **MA** | PST-top, APT-top | 82.9 | 253 | 887 | $1.8 \cdot 10^{-4}$ | 21.8 | n/a[3] |
| **FA** | [4] | 390 | 1,209 | 4,247 | $8.5 \cdot 10^{-4}$ | 104 | n/a[3] |
| **Sn** | APT-top | 0 | 1,561 | 5,484 | 0.53 | 135 | 3.1 |
| **Pb** | [4] | 2,331 | 4,911 | 17,250 | 0.10 | 423 | 0.48 |
| **Br** | PST-top, APT-bottom | 620 | 2,490 | 8,747 | 0.58 | 215 | n/a[3] |
| **I** | [4] | 3,298 | 10,974 | 38,548 | 36.5 | 946 | 15.3 |
| **Opaque back electrodes** | | | | | | | |
| **Au** | | 6,433 | | 2,2598 | 195 | 479 | 904 |
| **Ag** | | 3,496 | | 12,282 | 13.2 | 260 | 52.1 |
| **Cu** | | 2,987 | | 10,491 | 0.012 | 222 | 0.026 |
| **Al** | | 900 | | 3,161 | 0.0014 | 67.0 | $8.9 \cdot 10^{-5}$ |
| **Ni** | | 2,969 | | 10,430 | 0.11 | 221 | 0.028 |
| **Mo** | | 3,427 | | 12,037 | 1.2 | 255 | 1.42 |
| **W** | | 6,417 | | 22,539 | 7.7 | 478 | 14.1 |
| **Graphite** | | 143,733 | | 504,879 | 13.1 | 10,703 | 3.3 |
| **Transparent front electrodes** | | | | | | | |
| **In** | IZO (for PST) | 1,875 | n/a | 1,501 | 194 | 20.2 | 107 |
| **In** | IO:H (for APT) | n/a | 4,419 | 15,523 | 457 | 340.1 | 1809 |
| **Zn** | AZO (for PST) | 1,502 | n/a | 1,203 | 0.012 | 16.1 | 0.13 |
| **Sn** | FTO (for APT) | n/a | 10,036 | 35,254 | 3.39 | 772.4 | 18.0 |
| **Interconnection layers** | | | | | | | |
| **In** | ITO (for PST) | 384 | n/a | 308 | 39.7 | 4.1 | 22 |
| **In** | ITO (for APT) | n/a | 288 | 1,012 | 29.8 | 22.2 | 118 |
| **Zn** | AZO | 376 | | 1319.3 | 0.0030 | 28.9 | 0.23 |
| **Au** | nanolayer | 64.3 | | 226.0 | 1.95 | 5.0 | 9.3 |
| **Contact layers** | | | | | | | |
| **Ti** | c-$TiO_2$ & m-$TiO_2$ | 972 | 1,944 | 6,827 | 0.049 | 150 | 0.041 |
| **Sn** | $SnO_2$ | 456 | 912 | 3,205 | 0.31 | 70.2 | 1.6 |
| **Ni** | NiO | 132 | 264 | 464 | $1.0 \cdot 10^{-4}$ | 10.2 | 0.39 |
| **Cr** | blocking layer | 238 | | 836 | $5.3 \cdot 10^{-4}$ | 17.7 | 0.0031 |
| **Li** | LiF | 2.4 | | 8.27 | 0.0027 | 0.18 | $8.6 \cdot 10^{-4}$ |

[1] Highest projected material annual demand in the PV capacity model.

[2] Cumulative module installations, assuming a lifetime of 25 years.

[3] Unknown reserves.

[4] All three considered perovskite compositions.

**Table S 2 | Demand and supply criticalities of synthetic materials.** Listed are solid materials as well as solvents for perovskite and contact layers. For solid materials, units are t, for solvents $m^3$. DMF – dimethylformamide, DMSO – dimethyl sulfoxide, NMP – N-methylpyrrolidone, GBL – γ-butyrolactone, ACN – acetonitrile, EtOH – ethanol, THF – tetrahydrofuran, CB – chlorobenzene, Tol – toluene, IPA – isopropyl alcohol, SAM – self-assembled monolayers.

| **Substance** | | **Demand for 1 $TW_p$/a all-perovskite tandem PV [$t/TW_p$ or $m^3/TW_p$]** | **DPR [%]** | **Max. demand [t or $m^3$]** | **Ratio max. demand to annual production [%]** |
|---|---|---|---|---|---|
| **Solid materials** | | | | | |
| Spiro-OMeTAD | | 1,428 | $1.83\cdot10^{7}$ | 5,016 | $6.42\cdot10^{7}$ |
| PTAA | | 67 | $8.53\cdot10^{5}$ | 234 | $3.00\cdot10^{6}$ |
| P3HT | | 1,320 | $1.69\cdot10^{5}$ | 4,637 | $5.93\cdot10^{7}$ |
| SAM | | 6.7 | $8.53E\cdot10^{5}$ | | |
| PEDOT:PSS | | 200 | 200 | 703 | 703 |
| C60 | | 220 | 440 | 773 | $1.29\cdot10^{5}$ |
| $TiO_2$ nanoparticles | | 1,268 | 42 | 4,453 | 148 |
| **Solvents for perovskites** | | | | | |
| DMF | | $2.33\cdot10^{4}$ | 17.6 | $8.20\cdot10^{4}$ | 61.9 |
| DMSO | | $2.33\cdot10^{4}$ | 19.9 | $8.20\cdot10^{4}$ | 69.9 |
| NMP | | $2.33\cdot10^{4}$ | 16.0 | $8.20\cdot10^{4}$ | 56.1 |
| ACN | | $2.33\cdot10^{4}$ | 12.9 | $8.20\cdot10^{4}$ | 45.5 |
| THF | | $2.33\cdot10^{4}$ | 10.4 | $8.20\cdot10^{4}$ | 36.4 |
| GLB | | $2.33\cdot10^{4}$ | 1.9 | $8.20\cdot10^{4}$ | 6.6 |
| **Solvents for contact layers** | | | | | |
| | Used for | | | | |
| CB | PCBM | $2.13\cdot10^{4}$ | 0.033 | $7.49\cdot10^{4}$ | 0.12 |
| CB | spiro-OMeTAD | $1.57\cdot10^{4}$ | 0.025 | $5.52\cdot10^{4}$ | 0.086 |
| Tol | PTAA | $3.33\cdot10^{4}$ | 0.096 | 1.17·105 | 0.34 |
| EtOH | SAM | $1.78\cdot10^{4}$ | 0.036 | $6.24\cdot10^{4}$ | 0.13 |
| IPA | SAM | $1.78\cdot10^{4}$ | 6.32 | $6.24\cdot10^{4}$ | 22.2 |

## B. Analysis of literature on perovskite PV and sustainability

A literature meta-analysis was carried out using "Web of Science" (Clarivate) with the search phrases listed in Table S 3. The search was limited to search words occurring in the title as an indicator of the core focus of each publication. A search for "perovskite AND (solar OR photovoltaic*)" in all fields (including title, abstract and publication text) yielded a total of 5,003 publications on perovskite PV in 2022. The search was carried out on January 6$^{th}$, 2023.
The results are listed in Table S 4.

**Table S 3 | Search phrases used for the literature meta-analysis.**

| Category | Search phrase |
|---|---|
| Power conversion efficiency | perovskite AND (solar OR photovoltaic*) AND (efficien* OR PCE) |
| Stability | perovskite AND (solar OR photovoltaic*) AND (stability OR stable OR degradation OR lifetime) |
| Cost | perovskite AND (solar OR photovoltaic*) AND (cost OR price OR scalab* OR cheap) |
| Sustainability | perovskite AND (solar OR photovoltaic*) AND (sustainab* OR environm*) |
| Resource | perovskite AND (solar OR photovoltaic*) AND (resourc* OR abundan*) |

**Table S 4 | Number of publications identified for the search phrase specified in Table S 3.**

| Topic / Year | Efficiency | Stability | Cost | Sustainability | Resources |
|---|---|---|---|---|---|
| 2012 | 1 | 0 | 0 | 0 | 0 |
| 2013 | 9 | 0 | 1 | 0 | 0 |
| 2014 | 92 | 14 | 2 | 1 | 1 |
| 2015 | 246 | 69 | 14 | 12 | 0 |
| 2016 | 389 | 197 | 18 | 8 | 0 |
| 2017 | 516 | 283 | 36 | 16 | 0 |
| 2018 | 692 | 422 | 54 | 16 | 1 |
| 2019 | 852 | 558 | 56 | 18 | 0 |
| 2020 | 938 | 612 | 56 | 25 | 4 |
| 2021 | 1061 | 746 | 73 | 25 | 2 |
| 2022 | 1027 | 652 | 57 | 34 | 3 |

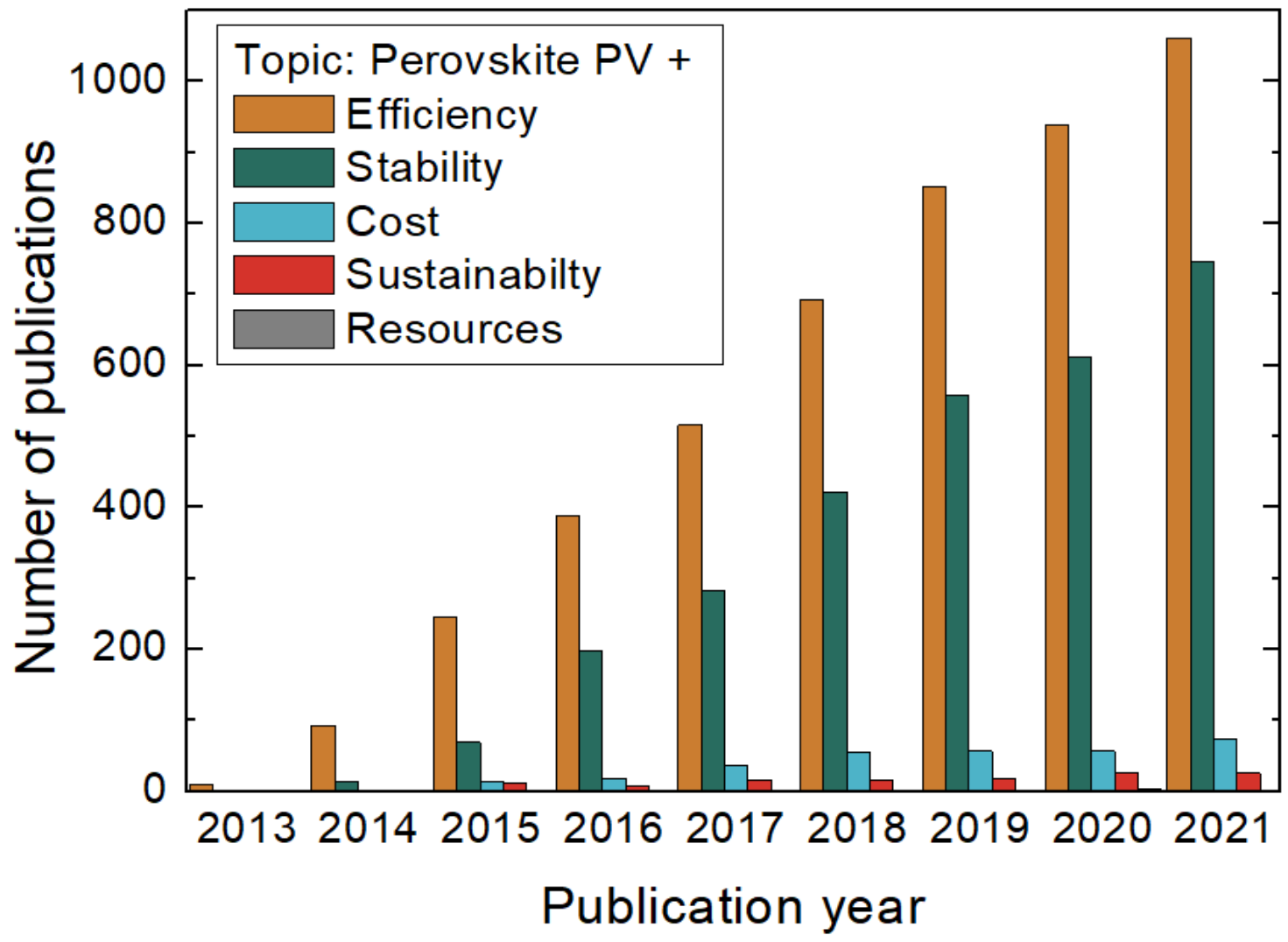

**Figure S 1 | History of publications on perovskite PV assessing issues of efficiency, stability, cost, sustainability, and resources. Note that number of annual publications on "resources" are ≤ 4 and hence not visible in the graph**

## C. Further details on the perovskite PV industry growth model

**Module efficiency improvement.** The fundamental physical power conversion efficiencies (PCE) limit for silicon solar cells lies at 29.4%.[1] For laboratory-scale solar cells, this value is already closely approached by record efficiencies of 26.7%.[2] It remains to be seen how close to the physical limit solar cells can be engineered at economically reasonable efforts. In any case, for PV installations, efficiencies not of solar cells but of commercial modules are important. On laboratory scale, the highest efficiency values for PV modules are currently 24.4%[2] and 22.4% for the best commercial PV modules.[3] The ITRPV2022 projects that efficiencies of commercial silicon single-junction modules will not surpass 24% by 2032,[4] highlighting that performance improvements in single-junction PV are stagnating. The current device performances imply a relative efficiency loss of at least 8.6% comparing laboratory records of solar cells with those of modules and 16.1% when comparing with commercial modules. If these loss values are applied to the fundamental efficiency limit for silicon solar cells, the resulting module efficiencies are 26.9% and 24.7.%, respectively. In our scenario, these module efficiency needs to be surpassed in 2036 or 2029, respectively, by the introduction of tandem PV technologies.

**Technological readiness of tandem PV technologies.** A range of tandem PV technologies are currently under investigation. For these technologies to become economically successful, a levelized cost of electricity (LCOE) below that of established single-junction devices needs to be achieved at market entry. This implies, first, that the efficiency needs to exceed those of commercial single-junction modules, a criterion which is currently fulfilled by III-V, III-V/Si, and perovskite/Si tandem devices.[5] Secondly, the additional cost needs to remain at an economic level. At the current state, III-V PV modules are too expensive due to the high costs for large-scale epitaxial crystal growth.[6] Due to the potential to reach highest efficiencies, however, future progress in fabrication technologies might render III-V based tandem devices cost competitive.[6,7] Perovskite PV, on the other hand, can be deposited with mature industrial coating techniques which promises large-scale high-throughput fabrication at lowest LCOE, making perovskite tandems to date the most promising tandem PV technology.[8–10] Consequently, in this study we focus on perovskite-based tandems.

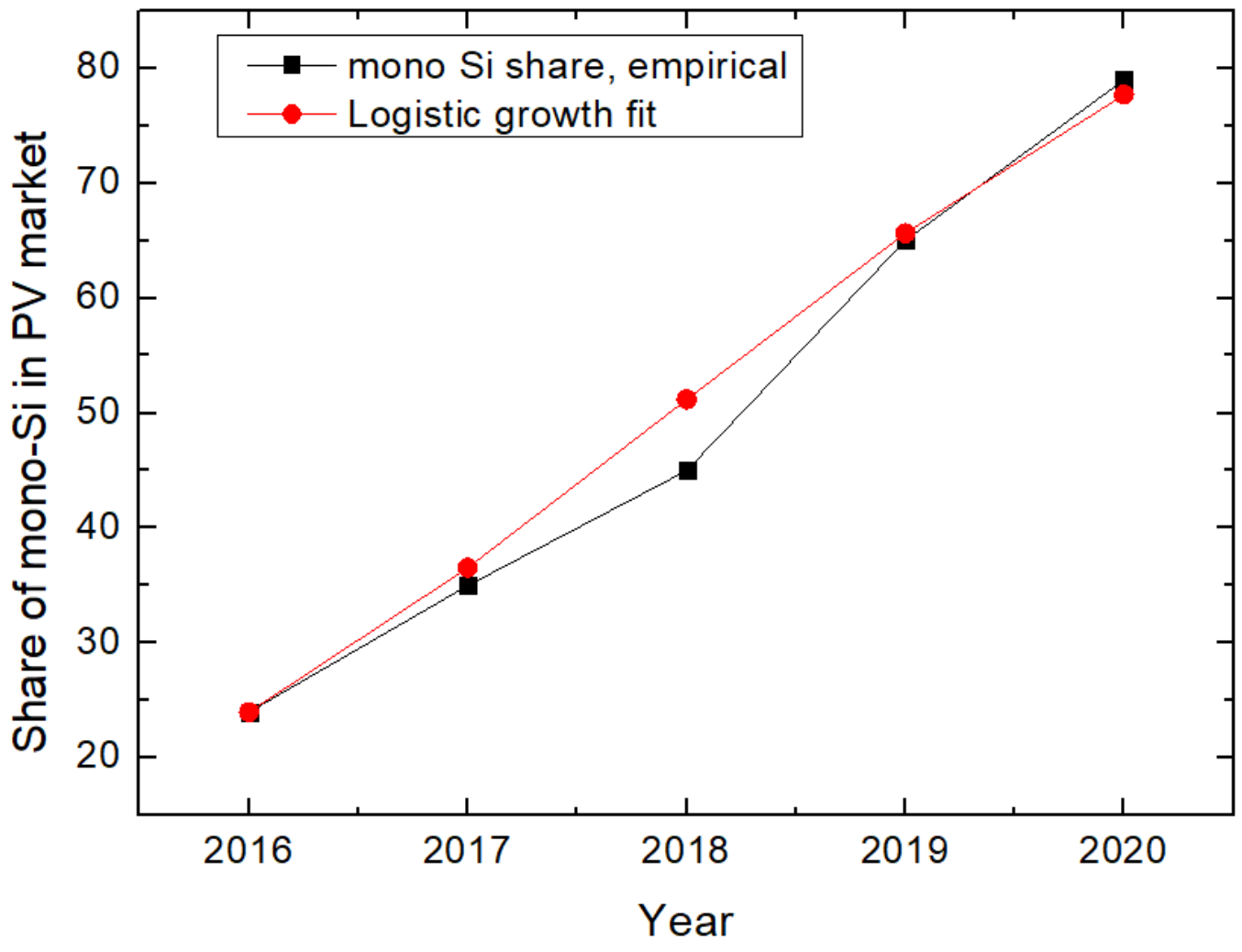

**Figure S 2 | Historic data of the share of monocrystalline silicon PV to the total PV market (black) and a logarithmic fit with *k* = 0.6. Historical data was obtained from ref. [3]**

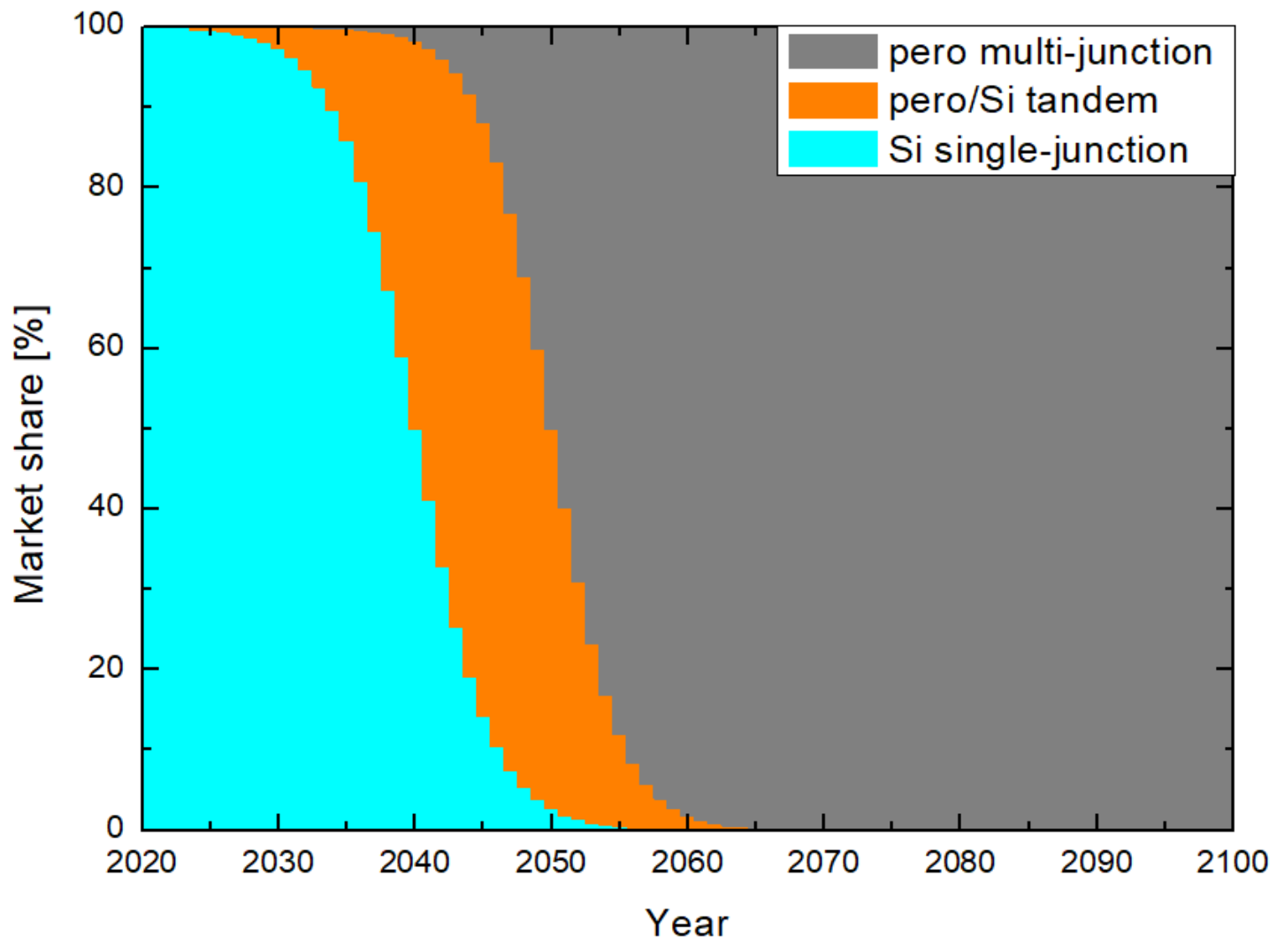

**Figure S 3 | Modeled market share of different PV technologies in the considered expansion model.**

## D. Materials and layer inventory

**Perovskites.** A range of perovskites have been investigated for PV applications. Of these, today the by far highest PCE, stabilities and technological readiness has been achieved with lead and tin halide perovskites. While lead halide perovskite solar cells (PSC) reach the highest efficiencies, their band gaps can only be tuned to a minimum of 1.51 eV.[11] For all perovskite tandems (APT), perovskites with lower band gaps in the bottom cell are necessary,[12] for which the most promising candidates are tin-lead halide perovskites.[13,14] The band-gap can be further fine-tuned by the stoichiometry of the halides (I or Br), or the monovalent cations.[11] Three monovalent cations form stable lead or tin halide perovskites, namely methylammonium (MA), formamidinium (FA) and cesium (Cs).[15]
This reduces the material inventory to six elements/molecules (MA, FA, Cs, Pb, Sn, I, Br), which can be mixed in infinite stoichiometries. To select a specific blend, in this study we focus on the stoichiometries of world record devices, namely on $Cs_{0.05}(FA_{0.77}MA_{0.23})_{0.95}Pb(I_{0.77}Br_{0.23})_3$ as top cell for PST (abbreviated PST-top),[16] and on $FA_{0.7}MA_{0.3}Pb_{0.5}Sn_{0.5}I_3$ (APT-top) and $FA_{0.8}Cs_{0.2}Pb(I_{0.62}Br_{0.38})_3$ (APT-bottom) for the top and bottom solar cell, respectively, of APT.[17]

**Opaque back electrodes.** For the back electrodes, where transparency is typically not required, opaque electrodes provide the highest sheet conductivity. Typically, metals with high conductivity are used for this purpose.[18] Gold and silver electrodes dominate academic research. Less expensive conductive metals such as aluminum (Al)[19] and copper (Cu)[20] have been implemented in PSC with the perspective of scalability, the latter being successfully used in the most efficient APT devices.[17,21] Less commonly, nickel (Ni),[22,23] molybdenum (Mo)[23,24] and tungsten (W)[23] have also been employed. Graphite based electrodes have emerged as an alternative to metal electrodes as they can be printed and are promising to yield higher device stabilities.[25]
Placing a metal layer in the vicinity of a halide-containing ionic compound such as the perovskite layer is a challenging endeavor. For the metals Au,[26] Ag[27] and Cu[28] (and potentially also for metallic Ni[29]), detrimental ion interdiffusion or chemical reactions with halide ions of the perovskite have been observed. However, this might not be a fundamental hurdle to achieve sufficient long-term stabilities as a successful prevention of ion intermigration has been demonstrated by the introduction of buffer layers[26,28,30,31] such as a 10 nm thin Cr layer. In contrast to the metal electrodes of typically 100 nm thickness, graphite is printed in layers of approximately 30 µm which signfficantly increases the material demand.

**Contact layers.** Regarding the contact layers, we distinguish between electron transport layers (ETL) and hole transport layers (HTL) which can in turn be categorized into inorganic and organic materials. Inorganic ETL materials mainly consist of metal oxides with the most important being mesoscopic (nanoparticulate) or compact $TiO_2$ as well as $SnO_2$.[32] Fullerene ($C_{60}$)[33] and fullerene derivatives such as PCBM[34] are common, especially in inverted PSC architectures. The material inventory of HTL is much more diverse. Among the inorganic HTL, NiO appears to be the most popular oxide. A range of copper compounds have also been employed such as CuO, $CuO_2$, CuI, or CuSCN.[35,36] Due to the large variety, Cu compounds as HTL are not explicitly assessed in this study. However, the low criticality of Cu compounds can be inferred from the finding that the use for Cu as metal electrode, which is associated with even higher Cu demand, was not found to be critical
For organic HTL, conjugated polymers, small molecules, or surface functionalizing agents like self-assembled monolayers molecules (SAM) are frequently used:[35] Common polymeric HTL are PTAA (poly(triarylamine)), P3HT (poly(3-hexylthiophene-2,5-diyl)), and PEDOT:PSS (poly(3,4-ethylenedioxythiophene):poly(styrenesulfonic acid)). Spiro-OMeTAD (2,2′,7,7′ tetrakis(N,N di p methoxyphenylamine) 9,9′ spirobifluorene) is the most commonly used small molecule. Recently, the introduction of SAMs like 2PACz, MeO-2PACz or Me-4PACz have yielded record-breaking PCE in PST.[16] Finally, also thin buffer layers such as LiF or Cr are used.[34,37]
Table S 5 shows the results of a literature research using the Perovskite Database[38] to assess the percentage of contact layers used in perovskite research. This database covers over 16,000 papers on PSC and can therefore be regarded as representative of the research field.

For the interpretation of this data, it is important to note that only single-junction devices are gathered in this database Database and that the database only comprises papers published until February 2020, i.e., very recent materials like SAM are underrepresented.
One can see in the table that over 50 % of the devices utilized $TiO_2$ as ETL and spiro-OMeTAD as HTL.

**Table S 5 | Percentage of electron transport layer (ETL) and hole transport layer HTL used in the perovskite solar cells as published in the Perovskite Database.**

| | Materials | Percentage of devices | |
|---|---|---|---|
| | | All devices | Devices with PCE > 20% |
| ETL | $TiO_2$, total | 53.9 | 46.7 |
| | $SnO_2$ | 10.4 | 34.5 |
| | $TiO_2$+$SnO_2$ | 0.8 | 2.8 |
| | Devices without $TiO_2$ or $SnO_2$ | 36.5 | 21.7 |
| | PCBM | 25.6 | 15.4 |
| | $C_{60}$ | 9.4 | 12.6 |
| | ZnO | 5.0 | 2.6 |
| HTL | spiro-OMeTAD | 50.8 | 70.6 |
| | PEDOT:PSS | 16.9 | 1.7 |
| | NiO | 6.5 | 6.5 |
| | PTAA | 5.1 | 13.2 |
| | P3HT | 2.4 | 0.2 |

**Transparent electrodes and interconnection layers.** As transparent conductive oxide (TCO) for transparent electrodes, tin-doped indium oxide (ITO) is already industrially employed for display manufacturing. ITO is also used for interconnection layers. The layers are composed of 90-95% of $In_2O_3$ doped with 5-10% $SnO_2$. Other indium based TCOs are hydrogenated indium oxide (IO:H) or indium zinc oxide (IZO), which typically consists of 90% $In_2O_3$ and 10% ZnO. Indium-free TCOs are fluorine-doped tin oxide (FTO), which can however not be coated at low temperatures compatible with perovskites or aluminum doped zinc oxide (AZO) which has successfully been deposited in APT as interconnection layer[39] and as electrode[40]. Alternatively, transparent electrodes can be realized by PEDOT:PSS or by lamination approaches.[41]
For interconnection layers, ultra-thin (~1 nm) layers of metallic nano clusters such as gold have also been employed.[17,21]
High performances can also be reached using heterojunctions of electron and hole selective layers as interconnection layers.[42] In our analysis, this option is covered under contact layers.

**Table S 6 | Layer thicknesses used in this study.**

| Layer | Layer thickness [nm] | Reference |
|---|---|---|
| **Transparent front electrode** | | |
| IO:H | 230 | [21] |
| AZO | 100 | [40] |
| IZO | 100 | [16] |
| FTO | 550 | Estimated from SEM-image published in [43] |
| **Interconnection** | | |
| ITO | 20 | [16] |
| AZO | 25 | [39] |
| Au nanolayer | 1 | [17] |
| **Electron transport layer** | | |
| c-$TiO_2$ | 40 | [43,44] |
| m-$TiO_2$ | 150 | [43,44] |
| $SnO_2$ | 25 | [17] |
| $C_{60}$ | 20 | [17] |
| **Hole transport layer** | | |
| SAM | 1 | [45] |
| PTAA | 10 | [16] |
| spiro-OMeTAD | 210 | [46] |
| P3HT | | |
| PEDOT | 30 | [47] |
| $NiO_x$ | 7.5 | [29] |
| **Buffer layer** | | |
| Cr | 10 | [26] |
| LiF | 1 | [16] |
| **Electrode** | | |
| Au | 100 | [46] |
| Ag | 100 | [48] |
| Cu | 100 | [48] |
| Al | 100 | - |
| Ni | 100 | - |
| Mo | 100 | - |
| W | 100 | - |
| Graphite | 28000 | [49] |
| **Perovskite** | | |
| All-perovskite tandem, narrow bandgap ($FA_{0.7}MA_{0.3}Pb_{0.5}Sn_{0.5}I_3$) | 1200 | [17] |
| All-perovskite tandem, wide bandgap ($FA_{0.8}Cs_{0.2}Pb(I_{0.62}Br_{0.38})_3$) | 450 | Estimated from SEM-image published in [17] |
| Single junction ($Cs_{0.05}(FA_{0.77}MA_{0.23})_{0.95}Pb(I_{0.77}Br_{0.23})_3$) | 500 | Estimated from SEM-image published in [16] |

## E. Production and reserves of minerals

Table S 7 lists the estimates for current production and reserves used in this study. Note that the year 2019 is chosen as base year for production as in the years after 2020, the market was disturbed by the global Covid-19 pandemic. All data is based on the Mineral Commodity Summaries 2021 by U.S. Geological Survey (USGS)[50] with the exception of data on indium reserves which was obtained by the Critical Metals Handbook of the British Geological Survey.[51]

**Table S 7 | Production and reserves of relevant inorganic materials. Source: [50].**

| Material | Production 2019 [kt/a] | Reserves [kt] |
|---|---:|---:|
| **Perovskites** | | |
| Cs | 2.50E-2 | 200 |
| Pb | 4,720 | 88,000 |
| Sn | 296 | 4,300 |
| I | 30.1 | 6,200 |
| Br | 429 | unknown |
| **Electrodes** | | |
| Au | 3.3 | 53 |
| Ag | 26.5 | 500 |
| Al | 63,200 | 75,000,000 |
| Cu | 24,500 | 870,000 |
| Ni | 2,610 | 64,000 |
| Mo | 294 | 18,000 |
| W | 83.8 | 3,400 |
| Cr | 44,800 | 570,000 |
| Graphite | 1,100 | 320,000 |
| **Materials for TCO and contact layers (unless mentioned above)** | | |
| In | 0.968 | 18.8 |
| Zn | 12,700 | 250,000 |
| Sn | 296 | 4,300 |
| Ti | 200,000 | 8,400,000 |
| Li | 86 | 21,000 |

## F. Synthesis routes and estimates of supply for solid synthetic materials

Note that the literature on annual production of synthetic materials is scarce. For the present study, production was estimated as follows.

**Estimation of methylamine and formamidine production.** Methylamine is synthesized by reacting ammonia with methanol. The synthesis of formamidinium halides such as FAI is typically carried out by formamidine acetate. An efficient electrochemical synthesis, as shown in Figure S 4, has been suggested by the reaction of cyanamide and aqueous acetic acid.[52] In laboratory scale, this reaction currently has a yield of up to 87%. The second step of the reactions could totally transfer to FAI, however, additional recrystallization is necessary to improve the purity of FAI which in turn decreases the yield.

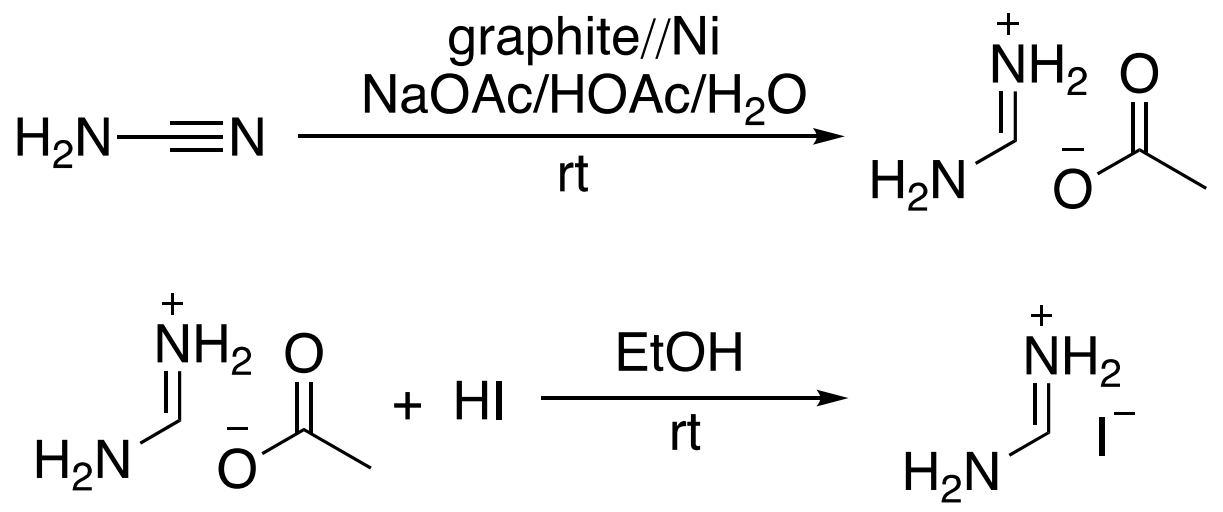

**Figure S 4 | Synthesis of formamidinium iodide (FAI).**

Table S 8 lists the annual production of educts needed for the synthesis of methylamine and formamidine. Note that except for ammonia, no peer reviewed or governmental data for the production could be identified. The lowest production of each educt was taken as an estimate for the maximum production of MA or FA, respectively.

**Table S 8 | Production of educts needed for the synthesis of methylamine and formamidine.**

| Material | Production in 2019 [kt/a] | Source |
|---|---:|---:|
| **Methanol** | 148,440 | [53] |
| **Ammonia** | 142,000 | [50] |
| **Cyanamide** | 1,291 | [54] |
| **Acetic acid** | 16,270 | [55] |

**Estimate of spiro-OMeTAD production.** We are not aware of any publicly accessible data on the global production of spiro-OMeTAD. To assess the annual production of spiro-OMeTAD, it was assumed that spiro-OMeTAD is exclusively produced for academic research on perovskite PV. Moreover, it was assumed that demand and consumption of spiro-OMeTAD is equivalent.

A Web of Science search for the search phrase "perovskite solar cell" yielded 4879 publications in 2021. Three scenarios for low, medium, and high estimates were considered (cf. Table S 9). The assessment also includes assumptions on how many substrates are fabricated per published paper. Following the literature search carried out in Table S 5, it was assumed that 50.8% of these substrates are coated with spiro-OMeTAD. Further assumptions consider the average spiro-OMeTAD precursor volume deposited per substrate and the corresponding precursor concentration. As shown in Table S 9, this results in spiro-OMeTAD consumption of 3.47, 7.81, and 22.32 kg/a for the scenarios low, medium, and high, respectively. For the data shown in the main manuscript, scenario "medium" was considered. Yet, even for scenario "high", there is a large discrepancy between current production and demand for 1 TWp/a perovskite PV.

In the calculations of the demand-production ration (cf. Extended Table 2) the number of 7.81 kg/a served as an estimate of the annual production of spiro-OMeTAD, PTAA and P3HT. Due to the lower use of SAMS, 1/10 of this number was assumed for the annual production.

**Table S 9 | Input parameters for the assessment of annual spiro consumption.**

| Scenario | low | medium | high |
|---|---|---|---|
| Substrates fabricated per publication | 400 | 700 | 1000 |
| Avg. precursor volume per substrate [µL] | 50 | 50 | 100 |
| Spiro precursor concentration [mg/ml] | 70 | 90 | 90 |
| **Spiro-OMeTAD consumed [kg/a]** | **3.47** | **7.81** | **22.32** |

**Discussion on catalyst supply criticality.** The synthesis of most organic hole transport materials (HTM) such as spiro-OMeTAD, PTAA, and P3HT require Pd-based catalysts. In 2019, the global Pd production was 227 t/a.[50] Today, 67% of the global Pd production is consumed by the automotive industry for combustion engines.[51] In a post-fossil world, this share may be used in the synthesis for perovskites. As industrial HTM production is not yet established, Pd usage during production as well as process losses are difficult to assess. If Pd recycling rates comparable to those of current industrial applications (80 to 90%) can be reached,[56] Pd will likely not be critical for HTM fabrication.

**Self-assembled monolayers. The synthesis route for 2PACz, a commonly used SAM material, is depicted in Figure S 5. The material demand for 1 $TW_p$/a APT production as well as DPR are listed in Table S 10.**

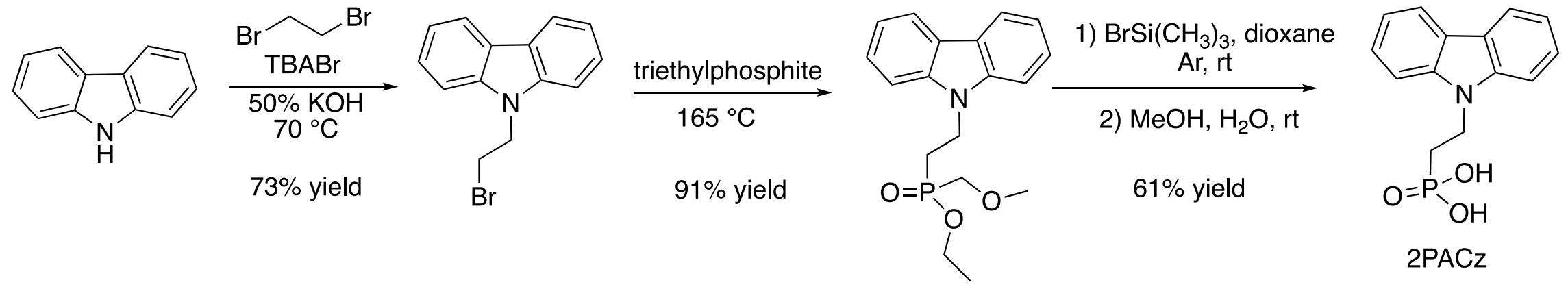

**Figure S 5 | Synthesis route of 2PACz according to [57].**

**Table S 10 | Material demand for the synthesis of 2PACz according to [57] to fulfil the material demand of 1 $TW_p$/a perovskite PV.**

| | | Material demand | Annual production | DPR [%] | Source, comments |
|---|---|---|---|---|---|
| **Step 1** | | | | | |
| **Educt** | 9H-carbazole | 99.8 t | unknown | unknown | |
| | 1,2-dibromoethane | 998 $m^3$ | 154 $m^3$* | 648 | [58] |
| | tetrabutylammonium bromide | 4.0 t | unknown | unknown | |
| | KOH | 179.7 t | 2,600 kt | $6.91\cdot10^{-3}$ | [59] |
| **Product** | 9-(2-bromoethyl)-9H-carbazole | 119.8 t | | | |
| **Step 2** | | | | | |
| **Educt** | 9-(2-bromoethyl)-9H-carbazole | 119.8 t | | | |
| | triethylphosphite | 1,471 $m^3$ | $3.08\cdot10^{7}$ $m^3$ | $4.78\cdot10^{-3}$ | See assessment below |
| **Product** | diethyl [2-(9H-carbazol-9-yl)ethyl]phosphonate | 131.4 t | | | |
| **Step 3** | | | | | |
| **Educt** | diethyl [2-(9H-carbazol-9-yl)ethyl]phosphonate | 131.4 t | | | |
| | 1,4-dioxane | 3,941 $m^3$ | 9,681 $m^3$ | 3.4 | Data for 1995, in metric tons. Source: [60] |
| | bromotrimethylsilane | 525 $m^3$ | | | |
| | methanol | 3,941 $m^3$ | 117,416 $m^3$ | 3.4 | [53] |
| | distilled water | 19,706 $m^3$ | abundant | n/a | |
| **Product** | **2-PACz** | **67.0 t** | | | |

* Twice the production of the USA in 1982 is assumed.[58] Note that global production is reported to have declined since the 1970s.[61]

Assessment of triethylphosphite production: a synthesis made from phosphorus trichloride and ethanol is assumed. The annual chlorine production in 2021 was 92.3 Mt.[62] From the annual production of phosphate rock ($P_2O_5$) 227 Mt/a in 2019[50] yields a maximum production of elemental phosphorous of 99.1 Mt/a. As the lowest value of educts determines the maximum production, this results in a $PCl_3$ maximum production of 30.8 Mt/a or $19.5\cdot10^{9}$ L/a. This is lower than the annual production of ethanol ($4,9\cdot10^{9}$ L) and hence can be regarded as maximum value for current production.

**PEDOT:PSS.**

The intrinsically conductive polymer polyethylenedioxythiophene-polystyrenesulfonate [PEDOT:PSS] is prepared by the oxidative polymerization of the monomer 3,4 ethylenedi-oxythiophene in the presence of PSS in water. The synthesis was established in 1990 and has been industrialized over the last 30 years.[63,64]

The oxidation results in a polyelectrolyte complex of the polyanion PSS and the polycation PEDOT. Approximately one positive charge is found for every three thiophene units for the oxidation of the PEDOT chain. The reaction scheme is shown in Figure S 6.

The conductive of PEDOT:PSS films can be adjusted between 10-6 to 103 S/cm. The solids content of PEDOT/PSS dispersion is typically in the range of 1-2%. A wide range of different dispersions is available from different vendors worldwide.

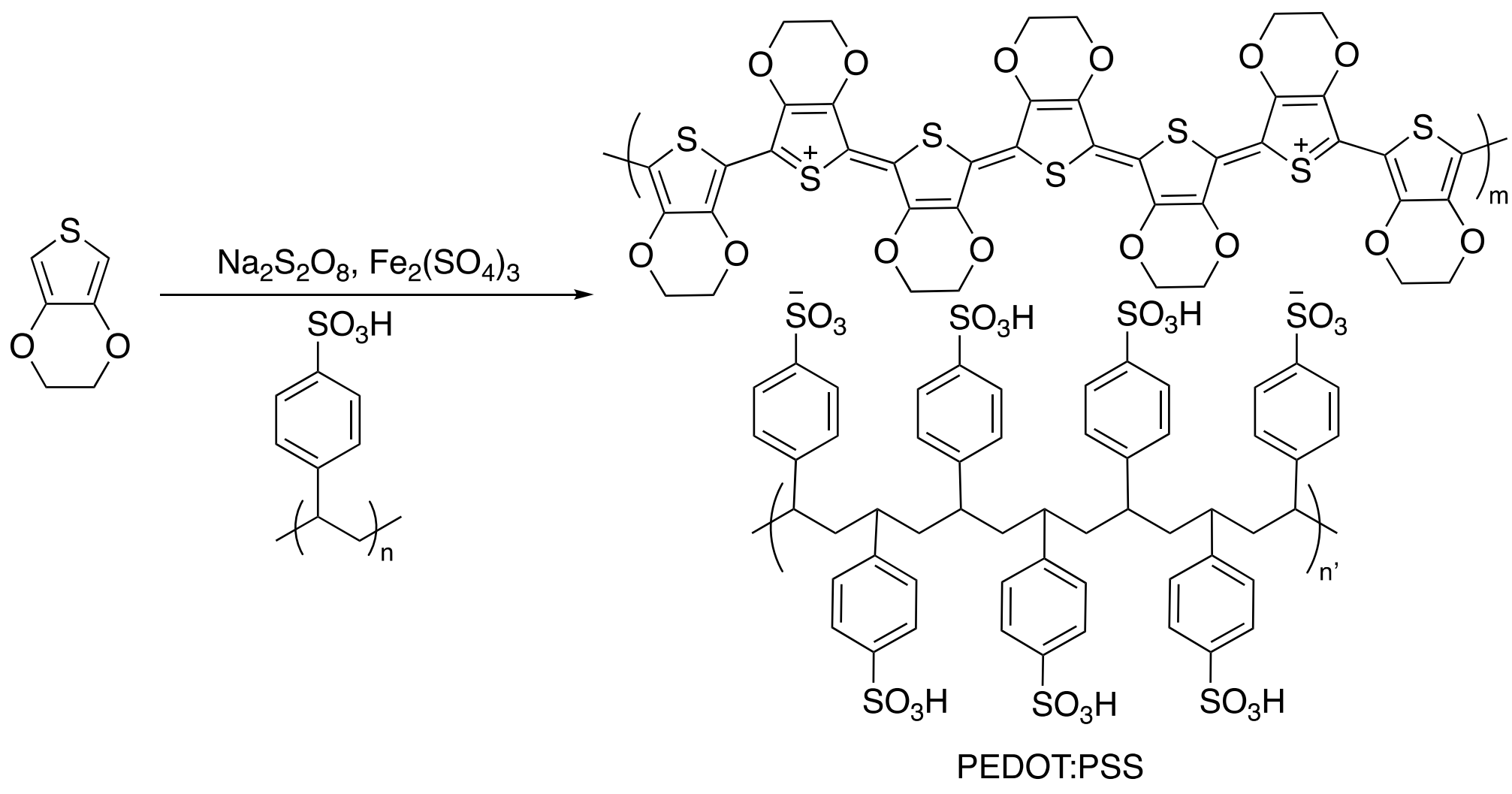

**Figure S 6 | Reaction scheme for the PEDOT Synthesis using sodiumperoxodisulfate as oxidant.**

**Fullerenes.** Peer-reviewed literature on the annual production of fullerenes and their derivatives is scarce. In 2012, Nowak's team estimated the global production of $C_{60}$ to 0.6 t/a and of nanoparticulate $TiO_2$ to 3 kt/a.[65] There has been a report by Frontier Carbon Corporation (a joint venture of Mitsubishi Corporation and Mitsubishi Chemical Corporation) claiming a fullerene production capacity of 40 t/a already in 2003.[66] (Note that production capacity is not equal to actual production.) In contrast, the first one-ton capacity production line in China was announced two decades later in 2021 by Guang'an Chenxin Chemical Co.[67] While these examples demonstrate that the data on annual fullerene production appears to be associated with high uncertainties, nonetheless this demonstrates a mature and competitive industrial material production. Accordingly, we assumed the technological readiness level (TRL) of fullerenes to the maximum of 9.

Several market outlooks on the fullerene market exist, reporting compound annual growth rates (CAGR) of the (monetary) market size between 5 and 7%. [68–71] To assess the annual global production from these figures, we need to ignore technological learning in the production capacity, i.e., we make the simplified assumption that CAGR is equivalent to the global production growth rate (in the following abbreviated as PGR). Thereby, using the numbers from Nowak et al. we arrive at annual production of 1.6 to 2.3 t/a, respectively, for the year 2022. Considering the production scale of China's largest plant, this might be an underestimation.

The market outlooks furthermore estimate the market volume to around 500 million USD in 2022. This number likely includes not only the market for raw fullerene materials but also further processed products like fullerene-containing cosmetics. Commercial prices for fullerene powder are typically listed with 30 USD/g but range as low as 1 USD/g (cf. Alibaba.com). In a rough estimate, we can assume a production price of 0.5 USD/g and a market volume of 100 million USD, corresponding to an annual production of 50 t.

Another perspective to look at the question of production is the energy requirement. Considering available data on current synthesis routes indicate a high energy intensity due to purification steps that require overall 12.7 GJ/kg.[72] For 1 $TW_p$ of all-perovskite tandem solar cells, this would translate in an energy demand of 2.6 PJ for the production of fullerene materials alone, which represents half of Germany's primary energy consumption.

In contrast to the above-mentioned prices for fullerene powders, cost analyses on perovskite PV stacks have identified fullerene-based layers as the among most expensive materials in PSC.[9]

Overall, despite a high degree of uncertainty, our estimates indicate that today's fullerene production may be in a reasonable order of magnitude to satisfy the demand of TW-scale perovskite PV, as the projected demand of 220 t/a for perovskite PV in 2050 could be satisfied with PGR of 5 to 7% and a production of 29 to 52 t/a in 2022.

**$TiO_2$ nanoparticles.** Layers of anatase titanium dioxide ($TiO_2$) nanoparticles are usually produced by scalable methods of spray pyrolysis and printing. First, the hole-blocking compact layer of $TiO_2$ (c-$TiO_2$) is deposited by spraying a precursor solution consisting of a titanium source, solution stabilizer and a solvent on a heated substrate. Considering that most publications report titanium isopropoxide, acetylacetonate and ethanol as Ti source, stabilizer and solvent, respectively, these materials were also considered for our analysis. The advantage of process simplicity, however, comes with a drawback of the required precision of substrate temperature and a need of cooling, raising the energy budget of this process applied on an industrial scale.[73] Numerous reports show that to manufacture high-efficiency devices, the c-$TiO_2$ layer has to be complemented by the presence of mesoporous $TiO_2$ (m-$TiO_2$) layer on top of it, for which nanoparticles of $TiO_2$ have to be produced first. Typically, this is done by the flame spray pyrolysis method, which is based on the hydrolysis of the $TiCl_4$ precursor vapor in an oxy-hydrogen flame and collection of the fine $TiO_2$ powder via a baghouse filter.[74,75] This technique has been used by industry since 1940s to produce fumed silica, alumina, pigmentary titania and titania for photocatalytic applications, such as the widely known Degussa (now Evonik) P25 powder of $TiO_2$ nanoparticles.[76,77] After the $TiO_2$ powder is obtained, it is combined with a thickener (e.g., ethyl cellulose), solvents (e.g., ethanol), dispersant (e.g., terpineol) and sometimes surfactants (e.g., triton x-100) to produce a $TiO_2$ paste, which after printing and sintering produces a mesoporous layer of $TiO_2$ nanoparticles.

It is noted that in the present study, the solvent consumption for $TiO_2$ layers as well as other printable layers such as graphite was not considered due to lack of representative data on both solvent demand and production of some solvents/additives. First assessments however indicate that this should be further investigated as there might be a supply criticality of terpineol.

In conclusion, Ti precursors needed for c-$TiO_2$ and m-$TiO_2$ are derived from $TiCl_4$. Today it is produced commercially via chlorination of titan-ferrous ores and/or slags by chlorine gas at temperature of 1,000-1,050°C according to the process outlined in Figure S 7.[78] First, ilmenite is heated up in the reactor together with coke and oxygen. Iron of the ilmenite forms iron chloride which has a low boiling point and therefore evaporates. After condensing and cooling it precipitates settling in the condenser. The rest of gases including $TiCl_4$ are passed on to a further tube condenser and kept at subzero temperatures to selectively condense $TiCl_4$. Further purification of the $TiCl_4$ can be achieved by additional distillation.[79]The global production of nanoparticulate $TiO_2$ was estimated to 3 kt/a in 2012.[65]It needs to be noted that the production of ilmenite involves carbon as reducing agent, currently resulting in additional formation of CO and $CO_2$.

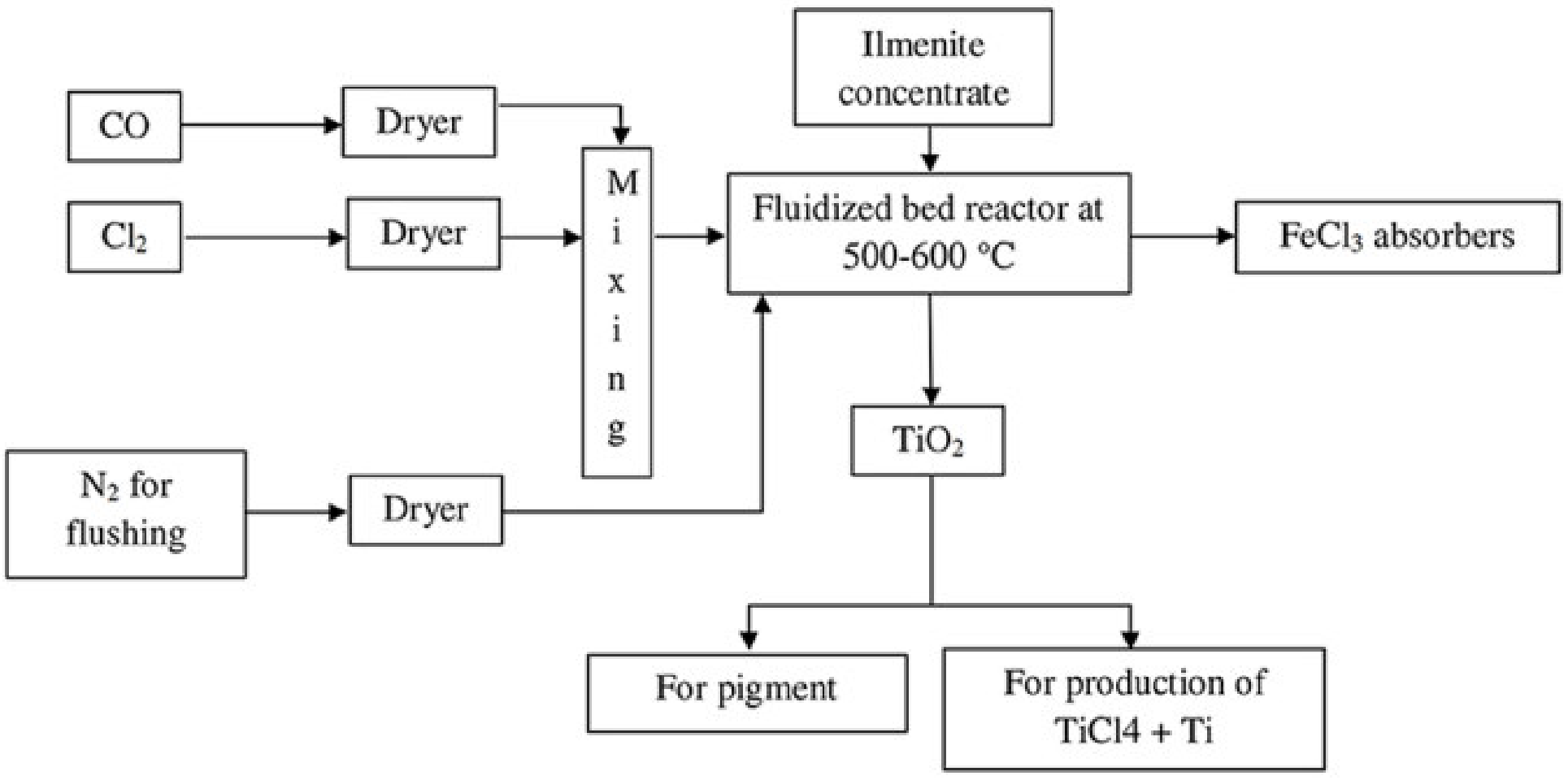

**Figure S 7 | Process of commercial $TiCl_4$ precursor production for highly crystalline $TiO_2$, as outlined in ref. [78].**

**NiO nanoparticles.** NiO nanoparticles can be formed from $NiCl_2 \cdot 6H_2O$ in aqueous NaOH solution. [80] $NiCl_2 \cdot 6H_2O$ can in turn be obtained from HCl with $Ni(OH)_2$, of which the latter can be gained from metallic Ni and NaOH. None of these chemicals are considered as critical in supply.

## G. Literature data on global solvent production

Table S 11 lists the annual solvents production used in the present study.

**Table S 11 | Current production of solvents needed for perovskite or contact layer deposition.**

| Solvent | Annual Production [t] | Source (comment) |
|---|---|---|
| Dimethylformamide | 125,000 | [81] (reference year: 1994) |
| Dimethyl sulfoxide | 129,000 | [82] |
| N-methylpyrrolidone | 150,000 | [83] |
| γ-butyrolactone | 1,404,000 | [84] |
| Acetonitrile | 142,000 | [85] |
| tetrahydrofuran | 200,000 | [86] |
| Ethanol | 38,661,000 | [87] |
| Chlorobenzene | 71,000,000 | [88] (values for benzene) |
| Toluene | 30,000,000 | [89] |
| Isopropyl alcohol | 221,000 | [90] |

## H. Sensitivity analysis of cesium demand

Table S 12 displays the Cs demand and respective demand-production-ratio (DPR) for the variation of the Cs cation content *x* in $FA_{1-x}Cs_xPb(I_{0.62}Br_{0.38})_3$ perovskite blends used in the top cells of all-perovskite tandem devices. Even for a Cs cation content of only 1%, the DPR reaches 57.7%.

**Table S 12 | Sensitivity analysis for the Cs content in $FA_{1-x}Cs_xPb(I_{0.62}Br_{0.38})_3$ perovskite blends of the top cells in all-perovskite tandem (APT) devices.** DPR – demand-production ratio

| Cs cation content *x* [%] | Cs demand for the production of 1 TW APT [t] | DPR [%] |
|---|---|---|
| 1 | 14.4 | 57.7 |
| 2 | 28.8 | 115.2 |
| 3 | 43.1 | 172.6 |
| 5 | 71.7 | 286.8 |
| 10 | 142.3 | 569.3 |
| 20 | 280.5 | 1121.8 |
| 50 | 671.5 | 2685.9 |

## I. Sensitivity analysis of indium demand

Table S 13 summarizes the dependence of the indium demand from the indium tin oxide (ITO) layer thickness.
As listed in Table S 6, typical indium-oxide thicknesses for all-perovskite tandem (APT) devices are 230 nm for transparent electrodes and 20 nm for interconnection layers.
Note that for transparent electrodes, a significant increase of the ITO sheet resistance is expected for thicknesses below 200 nm.[91] As the series resistance scales with the effective distance between the electrodes, this implies that the optimal ITO thickness depends on the size of the sub-cells of a module.

**Table S 13| Sensitivity analysis for the In content as function of ITO thickness.** DPR – demand-production ratio

| ITO thickness [nm] | In demand for 1 $TW_p$ APT [t] | DPR [%] |
|---|---|---|
| 5 | 96 | 9.9 |
| 10 | 192 | 19.9 |
| 20 | 384 | 39.7 |
| 30 | 576 | 59.6 |
| 50 | 961 | 99.3 |
| 100 | 1,922 | 198.5 |
| 200 | 3,843 | 397.0 |
| 300 | 5,765 | 595.5 |
| 400 | 7,686 | 794.0 |

## J. References